\documentclass{aa}  

\usepackage{times,epsfig,amsmath}
\usepackage{graphicx}
\usepackage{epstopdf}
\usepackage{natbib}
\bibpunct{(}{)}{;}{a}{}{,} % to follow the A&A style
\usepackage{pdflscape}
\usepackage{adjustbox}
\usepackage{subfigure}
\usepackage[flushleft]{threeparttable}
\usepackage{xcolor}
\usepackage{txfonts}
\usepackage{tabularx}
\usepackage{array}

\newcommand{\xmm}{{\it XMM-Newton}}

\usepackage{calrsfs}
\DeclareMathAlphabet{\pazocal}{OMS}{zplm}{m}{n}
\newcommand{\Mach}{\pazocal{M}}

\begin{document} 

\title{CHEX-MATE: Characterization of the intra-cluster medium temperature distribution}
\titlerunning{Temperature maps}
\author{L. Lovisari$^{1,2}$, S. Ettori$^{1,3}$,  E. Rasia$^{4,5}$, M. Gaspari$^{6}$, H. Bourdin$^{7,8}$, M. G. Campitiello$^{1,9}$,  M. Rossetti$^{10}$, I. Bartalucci$^{10}$, S. De Grandi$^{11}$, F. De Luca$^{7,8}$, M. De Petris$^{12}$, D. Eckert$^{13}$, W. Forman$^{2}$, F. Gastaldello$^{10}$, S. Ghizzardi$^{10}$,  C. Jones$^{2}$, S. Kay$^{14}$, J. Kim$^{15}$, B. J. Maughan$^{16}$,  P.  Mazzotta$^{7}$, E. Pointecouteau$^{17}$, G. W. Pratt$^{18}$, J. Sayers$^{15}$, M. Sereno$^{1,3}$, M. Simonte$^{19}$, P. Tozzi$^{20}$
}
\authorrunning{Lovisari et al.}
\institute{
INAF, Osservatorio di Astrofisica e Scienza dello Spazio, via Piero Gobetti 93/3, 40129 Bologna, Italy \and
Center for Astrophysics $|$ Harvard $\&$ Smithsonian, 60 Garden Street, Cambridge, MA 02138, USA \and
INFN, Sezione di Bologna, viale Berti Pichat 6/2, 40127 Bologna, Italy \and
INAF, Osservatorio Astronomico di Trieste, via Tiepolo 11, 34143 Trieste, Italy \and
Institute of Fundamental Physics of the Universe, via Beirut 2, 34151 Grignano, Trieste, Italy \and
Department of Astrophysical Sciences, Princeton University, Princeton, NJ 08544, USA \and 
Dipartimento di Fisica, Università di Roma `Tor Vergata', via della Ricerca Scientifica 1, 00133 Roma, Italy \and
INFN, Sezione di Roma `Tor Vergata', via della Ricerca Scientifica, 1, 00133, Roma, Italy \and
Dipartimento di Fisica e Astronomia, Università di Bologna, via Gobetti 92/3, 40121 Bologna, Italy \and
INAF, Istituto di Astrofisica Spaziale e Fisica Cosmica di Milano, via A. Corti 12, 20133 Milano, Italy \and
INAF, Osservatorio Astronomico di Brera, via E. Bianchi 46, 23807 Merate, Italy \and
Dipartimento di Fisica, Sapienza Università di Roma, Piazzale Aldo Moro, 00185 Roma, Italy \and
Department of Astronomy, University of Geneva, Ch. d’Ecogia 16, 1290 Versoix, Switzerland \and
Jodrell Bank Centre for Astrophysics, School of Physics and Astronomy, The University of Manchester, Manchester M13 9PL, UK \and
California Institute of Technology, 1200 E. California Blvd, MC367-17, Pasadena, CA 91125, USA \and
H. H. Wills Physics Laboratory, University of Bristol, Tyndall Ave, Bristol BS8 1TL, UK \and
IRAP, Université de Toulouse, CNRS, CNES, UT3-UPS, (Toulouse), France \and
Université Paris-Saclay, Université Paris Cité, CEA, CNRS, AIM, 91191, Gif-sur-Yvette, France \and Hamburger Sternwarte, University of Hamburg, Gojenbergsweg 112, 21029 Hamburg, Germany \and
INAF - Osservatorio Astrofisico di Arcetri, Largo E. Fermi, 50122 Firenze, Italy 
}
\mail{lorenzo.lovisari@inaf.it}

%\date{Received April 4, 2023; accepted October 29, 2023}

% \abstract{}{}{}{}{} 
% 5 {} token are mandatory
 
\abstract
{Galaxy clusters grow through the accretion of mass over cosmic time. Their observed properties are then shaped by how baryons distribute and energy is diffused. 
Thus, a better understanding of spatially resolved, projected thermodynamic properties of the  intra-cluster
medium (ICM) may provide 
a more consistent picture of how mass and energy act locally in shaping the X-ray observed quantities of these massive virialized or still collapsing structures.}
{We study the perturbations in the temperature (and density) distribution to evaluate and characterize the level of inhomogeneities and the related dynamical state of the ICM.
}
{We obtain and analyze the temperature and density distribution for 28 clusters ($2.4\times10^{14}M_{\odot}$<$M_{500}$<$1.2\times10^{15}M_{\odot}$; 0.07<$z$<0.45) selected from the CHEX-MATE sample. 
We use these spatially resolved two-dimensional distributions to measure the global and radial scatter and identify the regions that deviate the most from the average distribution. 
During this process, we introduce three dynamical state estimators and produce ``clean" temperature profiles after removing the most deviant regions.
}
{We find that the temperature distribution of most of the clusters is skewed towards high temperatures and is well described by a log-normal function. There is no indication that the number of regions deviating more than 1$\sigma$ from the azimuthal value is correlated with the dynamical state inferred from morphological estimators.
The removal of these regions leads to local temperature variations up to  10-20\% and an average increase of $\sim$5\% in the overall cluster temperatures. The measured relative intrinsic scatter within $R_{500}$, $\sigma_{T,int}/T$, has values of 0.17$^{+0.08}_{-0.05}$, and is almost independent
of the cluster mass and dynamical state. 
Comparing the scatter of temperature and density profiles to hydrodynamic simulations, we constrain the average Mach number regime of the sample to 
$\Mach_{3D}$=0.36$^{+0.16}_{-0.09}$. 
We infer the ratio between the energy in turbulence and the thermal energy, and translate this ratio in terms of a predicted  hydrostatic mass bias $b$, estimating an average value of $b\sim$0.11 (covering a range between 0 and 0.37)
within $R_{500}$.
}
{This study provides detailed temperature fluctuation measurements for 28 CHEX-MATE clusters which can be used to study turbulence, derive the mass bias, and make predictions on the scaling relation properties.  }


\maketitle

%
%-------------------------------------------------------------------

\section{Introduction}
Lying at the nodes of the cosmic web, massive clusters grow mainly through mergers of smaller mass units (groups and poor clusters) and through the continuous accretion of matter and field galaxies along filaments.
Their matter content reflects that of the Universe ($\sim$85\% dark matter, $\sim$15\% baryons), making them unique laboratories for testing models of gravitational structure formation and studying the thermodynamics of the intra-cluster medium (ICM).
The hot ($T\sim$10$^7$-10$^8$ K) tenuous plasma, which accounts for the majority ($\sim$85\%) of the baryonic content, is responsible for the X-ray light through thermal bremsstrahlung and line emission and has been the focus of many investigations since the launch of the first X-ray satellites. 

It is often assumed that, after the collapse of the main halo progenitor, the cluster gas settles in hydrostatic equilibrium  into a spherically symmetric potential well. Under this assumption, all  thermodynamic properties (e.g., temperature, density, and pressure) depend only on the distance from the center and are therefore homogeneous within a narrow radial shell. 
However, both X-ray observations and hydrodynamical simulations show that the gas is continuously perturbed by mergers and cosmological accretion (e.g., \citealt{Jones1999,1999ApJ...520L..21M,2007PhR...443....1M,chu12,vaz2012,vaz2017,simonte2022}).  In addition, turbulent motions due to a large variety of other astrophysical ICM processes (e.g., sloshing, Active Galactic Nuclei (AGN) feedback, thermal instabilities;  see, e.g., \citealt{Simionescu2019,gas20}) can also significantly alter the hydrostatic equilibrium at different scales, from the inner core to the cluster outskirts, potentially in an anisotropic manner. Bulk motions may also evolve into turbulence and there have been several observational evidence over the last few years (e.g.,  \citealt{Liu2015,Liu2016,Hitomi2016,Sanders2020})
Thus, the complex physics of galaxy clusters may result in a high level of temperature and density substructures (i.e., fluctuations) which are tied to the dynamical history of the clusters (e.g., \citealt{Simionescu2019,zuh2022}). 
The level of inhomogeneities in the ICM may also depend on the thermal conductivity and the viscosity of the gas (e.g., \citealt{Dolag2004,2006MNRAS.371.1025S,gas14,ZuHone2015}). Hence, a quantitative characterization of the thermodynamic structures in the ICM is needed before understanding a comprehensive picture of the physical  processes at work in galaxy clusters.

Perturbations can appear as isobaric (e.g., slow-motion regime, i.e., low Mach numbers, or gas cooling), adiabatic (sound waves and/or weak shocks), and isothermal (characterized by strong conduction or global perturbations of gravitational potential). 
The nature of the perturbations is often evaluated with the ``effective'' equation of state  ${\delta T}/T$=$(\gamma-1){\delta \rho} / \rho$ where $\gamma$=0  for isobaric perturbations (i.e., temperature and density fluctuations are anticorrelated), $\gamma$=5/3 for adiabatic perturbations  (i.e., density and temperature fluctuations are positively correlated); and $\gamma$=1 for isothermal perturbations (an unstable narrow regime).
Slow motions (i.e., $\Mach$<0.5) tend to be in pressure equilibrium with the medium, while stronger turbulence (i.e., 0.5<$\Mach$<1) overcomes the cluster stratification and is associated with an increased level of density fluctuations  compared to the level of temperature fluctuations (i.e., higher effective $\gamma$).
This behavior  is likely related to the relative importance of gravity waves and sound waves with the latter showing an increasing role with the radius  (e.g., \citealt{zhu13,gas14}).   
High-resolution simulations show that plasma perturbations are tightly related to the dynamical state of the cluster and that there is an interplay between different fluctuations (e.g., \citealt{gas13,gas14}). 

Regardless of their nature, all the turbulent processes are expected to generate fluctuations in ICM thermodynamic properties, that should be detectable in the related observables (see, e.g.,  \citealt{Simionescu2019} and references therein). 
Various theoretical works have suggested a strong link between turbulence and thermodynamic fluctuations (e.g., \citealt{zhu14,gas14,Mohapatra2020,Mohapatra2021,simonte2022,Zhuravleva2023}). 
From the observational point of view, \cite{sch04} were the first to investigate the relations between the thermodynamic fluctuations in pressure (as seen in projection from X-ray observations) and the turbulence in the ICM. 
More recent observational studies constrained the turbulence level by measuring the fluctuations in surface brightness or gas density in the inner regions of several galaxy clusters  (e.g., \citealt{chu12,Sanders2012,2016ApJ...818...14A,zhu16,Zhuravleva2018,2023MNRAS.518.2954D}) pointing to the isobaric nature of most  of the total variance of perturbations. 
However, we should note that most of these AGN-related studies are mainly focused on the inner regions of galaxy clusters. 
By comparing perturbations in the central regions and in the outer regions \cite{hof16} found hints for a change in the thermodynamic state from the isobaric to the adiabatic regime. 

The temperature structure can provide further information on the physics of shock-heated gas in merging events and on the role of turbulence and gas sloshing.
In fact, simulations by \cite{gas14} indicate that, in the low $\Mach$ regime, the fluctuations in temperature are as clear and robust tracers as density fluctuations, and even better than pressure fluctuations. 
With higher $\Mach$ numbers (e.g., $\Mach$>0.5) the amplitudes of the temperature fluctuations are a bit lower than the amplitudes of the other thermodynamic properties but are still a reasonably competitive tracer.

Thermodynamic 2D distributions can be used to identify cluster substructures and in turn to trace the merging process and the mass assembly history. 
By visual inspection of the thermodynamic maps of a large sample of clusters,  \cite{Tatiana2019} classified the clusters as relaxed or disturbed and correlated this classification with several standard cool-core diagnostic parameters. 
Their finding shows that the standard cool-core diagnostics are often too simplistic to account for the overall cluster dynamics because it is not rare that the cluster core can appear dynamically relaxed, while the outskirts can be dynamically unrelaxed. 
This is not surprising given that the relevant timescales in low density regions are much longer than in the core and that any denser infalling clump would be more prominent than in the inner regions.
The thermodynamic maps can reveal in detail the complex structure of a cluster. 
However, something more quantitative is now needed to link the morphology and the amplitude of the temperature substructures to the cluster dynamical state. 

The presence of substructures may induce biases in the determination of cluster properties, such as global gas temperature or total mass (and therefore impact the scaling relations; see \citealt{2009ApJ...699.1178Z}). 
For instance, a disturbance (e.g., an infalling structure or a cooler or hotter spot) in a cluster may appear as a peak and/or a discontinuity in the radial profile of the scatter of the fluctuations. 
Unrelaxed clusters may show strong fluctuations and significant correlations between, for example, temperature and gas density fluctuations. Thus, any diagnostic of substructures, asymmetries, and turbulence at any scale is expected to be directly linked to the scatter of the scaling relations, and to the bias in X-ray hydrostatic equilibrium (HE) masses.
We note that hydrodynamical simulations showed that structures in the ICM temperature distribution are, indeed, the main sources of systematic bias in HE mass estimates from X-ray data (see, e.g., \citealt{Rasia2006,Rasia2014,Ansarifard2020}). Thus, measuring the level of the ICM inhomogeneity and turbulent motions is crucial to estimate the hydrostatic mass bias (e.g., \citealt{ron13,Angelinelli2020}).

In this work, we study the temperature map of 28 clusters and investigate which kind of information can be extracted from their analysis. 
We compare them with the electron density maps with the same resolution (although in principle a significantly better resolution can be achieved) to map the inhomogeneities at the same scale.
The paper is organized as follows. 
In Sect. \ref{sect:analyisis}, we present the selected sample, the data available, and the spectral analysis performed to reconstruct the 2D distribution of the gas temperature. 
The results on the general properties of the temperature maps and the characteristics of the observed fluctuations are presented in Sect.~\ref{sect:res}.
The discussion of our main findings and the conclusions are described in Sect.~\ref{sect:dis} and \ref{sect:con}, respectively.

Throughout the paper, we assume a flat $\Lambda$CDM cosmology  with $\Omega_m=0.3$, $\Omega_\Lambda=0.7$, and $H_{0}=70$ km s$^{-1}$ Mpc$^{-1}$. All uncertainties are 1$\sigma$ confidence intervals unless stated otherwise.
We notice that several symbols are considered throughout the paper and to help the reader we provide a summary of the definitions in Table \ref{tab:definition}.

\begin{table}[t!]
\centering
\caption{Symbols definition.}
\label{tab:definition}
\begin{tabular}{ c p{7cm} } 
\hline
\hline
\noalign{\smallskip}
Symbol & Definition  \\
\noalign{\smallskip}
\hline
\noalign{\smallskip}
$\epsilon$ & statistical uncertainty (see Eq. \ref{eq:epsilon}) \\
$\sigma$ & scatter (or dispersion) around a mean value (see Eq. \ref{eq:fluc_sigma}) \\
$\delta$ & difference between two values \\
$s$ & difference between two values in units of $\epsilon$ (see Eq. \ref{eq:sigma-clip})\\
$\varsigma$ & standard deviations of a Gaussian distribution\\
\hline
\end{tabular}
\tablefoot{
When this value is associated with a single cell on the maps we add the subscript $i$ to these symbols, when it is associated to an annulus we add the subscript $j$, while if the subscript is omitted, the value refers to the global property (i.e., within $R_{500}$).
}
\end{table}

\section{Data analysis}\label{sect:analyisis}

\subsection{Sample}\label{sect:sample}
The Cluster HEritage project with {\it XMM-Newton} - Mass Assembly and Thermodynamics at the Endpoint of structure formation (CHEX-MATE\footnote{\url{http://xmm-heritage.oas.inaf.it/}}; \citealt{chexmate2021}) is a ground-breaking multi-wavelength investigation of 61 local clusters (Tier-1: 0.05 < $z$ < 0.2; 2$\times10^{14}M_{\odot}<M_{500}<9\times10^{14} M_{\odot}$) and 61 massive clusters (Tier-2: $z$<0.6 with $M_{500}>7.25\times10^{14}M_{\odot}$); four clusters belong to both Tiers. 
This project aims at covering, with homogenous {\it XMM-Newton} exposures, the ICM emission up to at least $R_{500}$, for this minimally biased, signal-to-noise-limited sample of objects detected by {\it Planck} through the Sunyaev-Zeldovich effect.   
One of the goals of the project is the robust  assessment of the level of any systematic error affecting the X-ray analysis of the cluster gas density, temperature, and mass profiles. 
Indeed, gas inhomogeneities prevent a robust determination of the global cluster properties and need to be properly understood to use clusters in astrophysical and cosmological studies.

\begin{figure}[t]
\centering
\includegraphics[width=.5\textwidth]{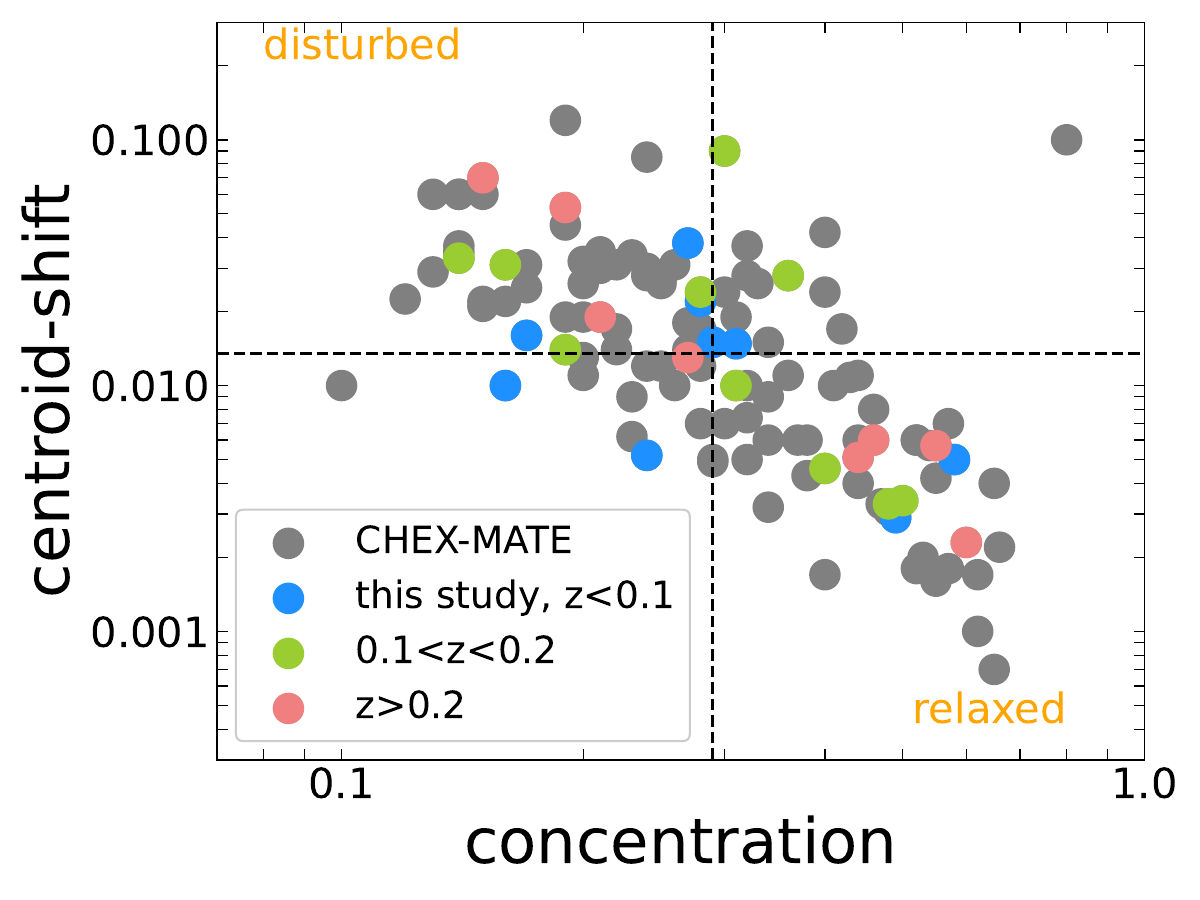}
\caption{
Distribution of the clusters in our sample in the concentration-centroid-shift (i.e., $c$-$w$) space. The values are taken from \cite{cam22}. The dashed lines indicate the median values. The most relaxed clusters are in the bottom-right corner while the most disturbed in the upper-left corner.  }
\label{fig:cw}
\end{figure}

We selected a pilot sample of 28 clusters (corresponding to $\sim$1/4 of the CHEX-MATE sample) where $R_{500}$ is completely covered by one single \xmm~pointing (i.e., simplifying the analysis and speeding up the fitting process) to (i) investigate how the thermodynamical maps can be used to investigate the fluctuations in the ICM and (ii) complement the information about the dynamical state of the clusters coming from the standard morphological analysis of the X-ray images presented in \cite{cam22}. 
For this reason, the sample has been selected to cover a large range of morphological properties as estimated in CHEX-MATE (see the distribution of the concentration, $c$, and centroid-shift, $w$, parameters in Fig.~\ref{fig:cw}). Although we did not attempt to be representative in terms of mass and redshift, the selected sample roughly covers the CHEX-MATE ranges.
The list of clusters is provided in Table~\ref{tab:clusters} and the gallery is presented in Appendix \ref{fig:gallery}.

\begin{table*}[t!]
\centering
\caption{Cluster properties.}
\label{tab:clusters}
\begin{tabular}{c c c c c c c c c c c c}
\hline
\hline
\noalign{\smallskip}
Index & Name & RA & DEC & $z$ & OBSID & $t_{\rm exp}$ & $R_{500}$ & $c$ & $w$ & $P_{20}$ & $M_{\rm all}$ \\
\noalign{\smallskip}
\hline
\noalign{\smallskip}
1  & G008.31--64.74 & 344.701 & -34.800 & 0.3120 & 0827010901 & 33.3, 35.6, 22.6 & 1237 & 0.21 & 0.019 & 33.00 & 0.82 \\	
2  & G008.94--81.22 &   3.580 & -30.392 & 0.3066 & 0743850101 & 95.8, 95.5, 71.0 & 1321 & 0.19 & 0.053 & 8.30 & 1.06 \\	
3  & G031.93+78.71  & 205.453 & 26.374  & 0.0724 & 0108460101 & 22.1, 22.4, 16.0 &  963 & 0.31 & 0.015 & 0.09 & -0.47 \\
4  & G041.45+29.10  & 259.439 & 19.679  & 0.1780 & 0601080101 & 55.5, 56.0, 38.7 & 1169 & 0.16 & 0.031 & 1.60 & 0.40 \\
5  & G042.81+56.61  & 230.622 & 27.707  & 0.0723	& 0202080201 & 19.5, 19.5, 12.6 & 1115 & 0.29 & 0.015 & 2.70 & -0.05 \\ 
6  & G046.88+56.48  & 231.033 & 29.881  & 0.1145 & 0827010601 & 28.9, 29.1, 18.7 & 1171 & 0.14 & 0.033 & 22.00 & 1.41  \\
7  & G049.32+44.37  & 245.127 & 29.893 & 0.0972	& 0692930901 & 11.2, 11.2, 8.7 & 1064 & 0.28 & 0.022 & 4.00 & 0.08 \\ 
8  & G055.59+31.85  & 260.613 & 32.132  & 0.2240 & 0693180901 & 27.8, 28.5, 21.3 & 1294  & 0.46 & 0.006 & 2.00 & -0.46 \\
9  & G056.77+36.32  & 255.677 & 34.061  & 0.0953 & 0740900101 & 25.4, 25.3, 19.0 & 1117 & 0.49 & 0.003 & 0.90 & -0.81 \\
10 & G057.61+34.93 & 257.455 & 34.453 & 0.0802 & 0827010501 & 20.4, 20.7, 16.1 & 1064  & 0.16 & 0.010 & 2.30 & 0.49 \\
11 & G057.92+27.64  & 266.060 & 32.991  & 0.0757 & 0827030301 & 19.4, 20.0, 11.7 &  955 & 0.58 & 0.005 & 2.60 & -0.49 \\ 
12 & G067.17+67.46  & 216.511 & 37.825  & 0.1712 & 0112230201 & 19.1, 20.3,  8.8 & 1285 & 0.48 & 0.003 & 0.16 & -1.29 \\  
13 & G077.90--26.63 & 330.220 & 20.971  & 0.1470 & 0827020101 & 26.1, 26.4, 18.3 & 1150 & 0.40 & 0.005 & 0.30 & -1.03 \\
14 & G080.41--33.24 & 336.526 & 17.364  & 0.1072 & 0762470101 & 73.5, 79.8, 40.9 & 1061 & 0.36 & 0.028 & 0.70 & -0.20 \\
15 & G083.86+85.09  & 196.463 & 30.894  & 0.1832 & 0827030701 & 28.2, 28.3, 21.4 & 1116 & 0.31 & 0.010 & 0.90 & -0.26\\ 
16 & G114.79--33.71 & 5.156 & 28.659 & 0.0940 & 0827320401 & 23.7, 23.6, 16.5 & 1068 & 0.24 & 0.005 & 5.00 & -0.07 \\
17 & G124.20--36.48 &  13.960 & 26.410  & 0.1971 & 0203220101 & 36.0, 36.4, 27.0 & 1280 & 0.30 & 0.090 & 120.00 & 1.60 \\ 
18 & G172.74+65.30  & 167.933 & 40.825  & 0.0794 & 0827031101 & 33.7, 33.9, 25.7 &  920 & 0.29 & 0.015 & 16.00 & 0.14 \\
19 & G208.80--30.67 & 73.528 & -10.219 & 0.2470 & 0603890101 & 70.6, 73.7, 63.0 & 1257 & 0.15 & 0.070 & 24.00 & 1.43\\
20 & G238.69+63.26  & 168.226 & 13.436  & 0.1690 & 0500760101 & 49.4, 49.2, 42.0 & 1075 & 0.28 & 0.024 & 3.00 & 0.34 \\ 
21 & G243.64+67.74  & 173.213 & 14.454  & 0.0834 & 0827010801 & 24.3, 24.0, 16.6 & 1055 & 0.27 & 0.038 & 1.40 & 0.01 \\
22 & G266.04--21.25 & 104.585 & -55.941 & 0.2965 & 0112980201 & 21.8, 21.8, 15.4 & 1479 & 0.27 & 0.013 & 1.60 & 0.02 \\ 
23 & G266.83+25.08  & 155.959 & -27.256 & 0.2542 & 0827011001 & 33.2, 34.0, 23.9 & 1254 & 0.55 & 0.006 & 1.00 & -1.12\\ 
24 & G273.59+63.27  & 180.104 &  3.347  & 0.1339 & 0827010301 & 27.9, 27.7, 21.5 & 1191 & 0.19 & 0.014 & 17.00 & 0.32 \\
25 & G287.46+81.12  & 190.323 & 18.574  & 0.0730 & 0149900301 & 14.7, 15.3, 10.9 &  943 & 0.17 & 0.016 & 2.60 & 0.34 \\
26 & G313.87--17.10 & 240.452 & -75.755 & 0.1530 & 0692932001 & 15.8, 17.4,  8.3 & 1335 & 0.50 & 0.003 & 0.30 & -0.54 \\
27 & G324.04+48.79  & 206.877 & -11.752 & 0.4516 & 0112960101 & 30.8, 31.2, 25.7 & 1319 & 0.60 & 0.002 & 0.37 & -1.35 \\ 	
28 & G349.46--59.95 & 342.185 & -44.530 & 0.3475 & 0504630101 & 25.4, 26.2, 18.2 & 1406 & 0.44 & 0.005 & 1.30 & -0.71 \\ 	
\noalign{\smallskip}
\hline
\end{tabular}
\tablefoot{
Cluster indices and names, X-ray peak coordinates, and redshifts are shown in columns 1--4. In columns 5--6 we provide the list of {\it XMM-Newton} observations that we investigated, and the corresponding clean exposure times (in ks) for MOS1, MOS2, and pn. The value of $R_{500}$ (in kpc) used to extract the maps provided in column 7 are taken from \cite{psz2}. In  columns 8--11, we provide the concentration, centroid–shift, power-ratio, and $M_{\rm all}$ parameters estimated by \cite{cam22}.
}
\end{table*}

\subsection{Data reduction}
Observation data files (ODFs) were retrieved from the {\it XMM-Newton} archive and reprocessed with the XMMSAS v19.0.0 with the latest calibration information available in October 2021. 
We used tasks {\it emchain} and {\it epchain} to generate calibrated event files from raw data.
We only considered single to quadruple  pixel MOS events (i.e., PATTERN$<$13), and single to double pixel pn events (i.e., PATTERN$<$5). 
Moreover, we considered only the high-quality MOS (i.e.,  \#XMMEA\_EM) and pn (i.e., FLAG==0) events. 
The data were cleaned for periods of high background due to the soft protons using the XMM-ESAS  tools {\it mos-filter} and {\it pn-filter}, respectively. Additionally, we also removed the CCDs in the so-called `anomalous state' (see \citealt{2008A&A...478..575K} for more details). 
Finally, we also obtain a list of out-of-time events, which are then subtracted from images and spectra after rescaling them based on the pn observation mode.

Point sources, identified by running the SAS wavelet detection tool {\it ewavelet}, were excluded from the analysis. 
To ensure a constant CXB flux across the cluster volume we excluded only the sources higher than a threshold value in the LogN-LogS distribution. More details can be found in \cite{Bartalucci2023}. 
We note that only point sources were masked, while clumps or large scale substructures were not removed. 
However, since the detection algorithm accounts for the increasing PSF at large radii we do not expect that point sources can be confused with clumps or large scale substructures in the outer regions of the clusters.

\subsection{Spectral fitting}
The modeling of the background was done following \cite{lov19} with a few changes that we highlight here. As in \cite{lov19} the cosmic X-ray background (CXB) was modeled by fitting simultaneously the {\it XMM-Newton} spectra with the ROSAT All-Sky Survey (RASS) spectra extracted from a region beyond the virial radius using the available tool\footnote{\url{https://heasarc.gsfc.nasa.gov/cgi-bin/Tools/xraybg/xraybg.pl}} (\citealt{2019ascl.soft04001S}) at the HEASARC webpage. 
However, we now use the counts-based spectrum, implemented with the v3.0.0, which allows us to use properly the {\it cstat} statistic during the fit. The non-vignetted quiescent particle background (QPB) was estimated using the filter wheel closed (FWC) observations after renormalizing them to match the observation of interest following the procedure presented in \cite{2009ApJ...699.1178Z}. 
In \cite{lov19}, the renormalizing values were obtained for each detector individually. 
However, \cite{marelli2021} showed that the corners (i.e., the out-of-field of view regions) of the pn detector are also exposed to photons and particles concentrated by the X-ray telescope, leading to wrong renormalization values. 
Luckily, the authors showed that there is a very tight and linear correlation between the particle backgrounds level detected in pn and MOS2 cameras. 
Therefore, we applied to pn the renormalization values determined for MOS2. \cite{marelli2021} showed that also some MOS out-of-field-of-view regions are exposed to celestial photons and provide the mask to exclude them from MOS2 data. 
Since the same mask cannot be applied to MOS1, we used MOS2 values as a proxy also to renormalize the MOS1 FWC datasets. 
Finally, we added an extra broken power-law (folded only with the RMF), to account for a residual soft proton contamination which affects many observations even after filtering the flare events (e.g., \citealt{lm08}).
The slopes are fixed to 0.4 (below 5 keV) and 0.8 (above 5 keV) as described in \citet{lm08}. Since this component may be different for MOS and pn detectors, the normalization is left free to vary in the three detectors and in all the regions  of interest (this accounts, in first approximation, for the proton vignetting).

To perform the spectral analysis of our sample, we use the XSPEC fitting package (\citealt{xspec}) v12.11.0k. 
We fit all our spectra with an APEC thermal plasma model (\citealt{Smith2001}) with an absorption fixed at the values provided by \cite{bourdin23}\footnote{For many clusters the used values are very close to the total $N_{H}$ values estimated by following \cite{wil13}. 
However, the latter seems to overestimate the true $N_{H}$ values in the low absorption regime.}. 
The best fits were obtained by minimizing the C-statistics (i.e., a modified Cash statistics; \citealt{Cash1979}) assuming the metal abundances provided by \cite{Asplund2009}. 
Our analysis slightly differs from the standard CHEX-MATE pipeline (described in a forthcoming work) which is partially using some ESAS tools (see \citealt{Snowden2008}) not optimized to work with a large number of regions, but returns consistent results.

\subsection{Temperature maps}\label{voronoi}
We determined the maps using the Weighted Voronoi Tesselation (WVT) method (a comparison with the results from the curvelet analysis, see \citealt{Bourdin2015}, is shown in Appendix \ref{App:curvelet}). 
The cells of the Voronoi map were generated using the WVT algorithm provided by \citet{vor06}, which is a generalization of the \citet{2003MNRAS.342..345C} Voronoi binning algorithm, and requiring a signal-to-noise ratio S/N$\sim$30 in the 0.3-7 keV band. 
As shown in \cite{lov19}, this S/N allows us to estimate the temperatures with a statistical uncertainty of 10-20\% (68\% c.l., see the relative error maps in Appendix~\ref{gallery}). 
This choice may impact some of our results, therefore, we also produced the maps with S/N$\sim$50 (i.e., larger cells and smaller statistical uncertainties) that we use for comparison purposes.
For each region of the map, we obtain a measure of the projected temperature and electron density. 
The projected density value for each cell of the map is given by
\begin{equation}
n_e=\sqrt{\frac{N 4\pi D_a^2 (1+z)^2}{f_{n_e/n_H}  10^{-14} V}},
\label{eq:n_k}
\end{equation}
where $N$ is the APEC normalization, $D_a$ the angular diameter distance, $f_{n_e/n_H}$ is the fraction between the electron and proton density and depends on the metallicity value, and $V$ the volume of the emitting region. 
The normalization $N$ is computed as a linear combination of the individual detectors' normalization, each weighted for the exposed detector area and the count rate contribution in the energy band of the spectral fit. 
The volume was determined as $V=2A\sqrt{R_{500}^2-X^2-Y^2}$, where $A$ is the area of the region, and $X$ and $Y$ are the projected distances in the east-west and north-south directions, and we assumed that the properties of the material in each region are homogeneous and that there is no other material projected onto them. 
In Appendix~\ref{gallery}, we show the recovered maps for all the clusters.

\subsection{Reconstruction of the 1D ICM profiles of temperature and density}

From the 2D distributions, we calculate the projected temperature and density profiles by properly weighting each spatial cell. 

The projected temperature in each annulus $j$ %(typically in steps of 0.1$R_{500}$) 
is computed as 
\begin{equation}\label{T1D}
T_{1D, j} = \frac{\sum{T_{2D, i} w_{i}}}{\sum{w_{i}}},
\end{equation}
where $i$ identifies the cell number in the 2D distributions, and $w_i$ are the weights that take into account both the area $A_{ij}$ covered by the cell $i$ in a given annulus $j$ and the emissivity (more details about the weighting are provided in Appendix \ref{weights})
\begin{equation}
w_i = A_{ij} S_{X, i} T_{2D, i}^{\alpha}, % \frac{A_{ij}}{\sum{A_{ij}}}  % n_{e, i}^2 T_{2D, i}^{-0.75}.
\end{equation}
with $S_{X, i}$ being the surface brighness in the cell $i$, and $\alpha = -0.75$ \citep[see, e.g.,][]{maz04}. The latter component of $w_i$ is needed because when the overall spectrum is given by the superposition of several single–temperature spectra (e.g., the different cells), the cooler gas components are relatively more important in determining the temperature resulting from the spectral fit because the shape of the \xmm~responses (e.g., \citealt{maz04,vik2006}). We note that such a value of $\alpha$ is strictly valid only in the cluster regime, and, if not included in the calculation, the recovered 1D temperatures tend to be overestimated (see Appendix \ref{weights}). In Fig. \ref{fig:T1DTprof} we show the comparison between the profiles recovered from the maps and the profiles obtained with the direct fitting of the spectra extracted for each annulus.

\begin{figure}[t]
\centering
\includegraphics[width=0.5\textwidth]{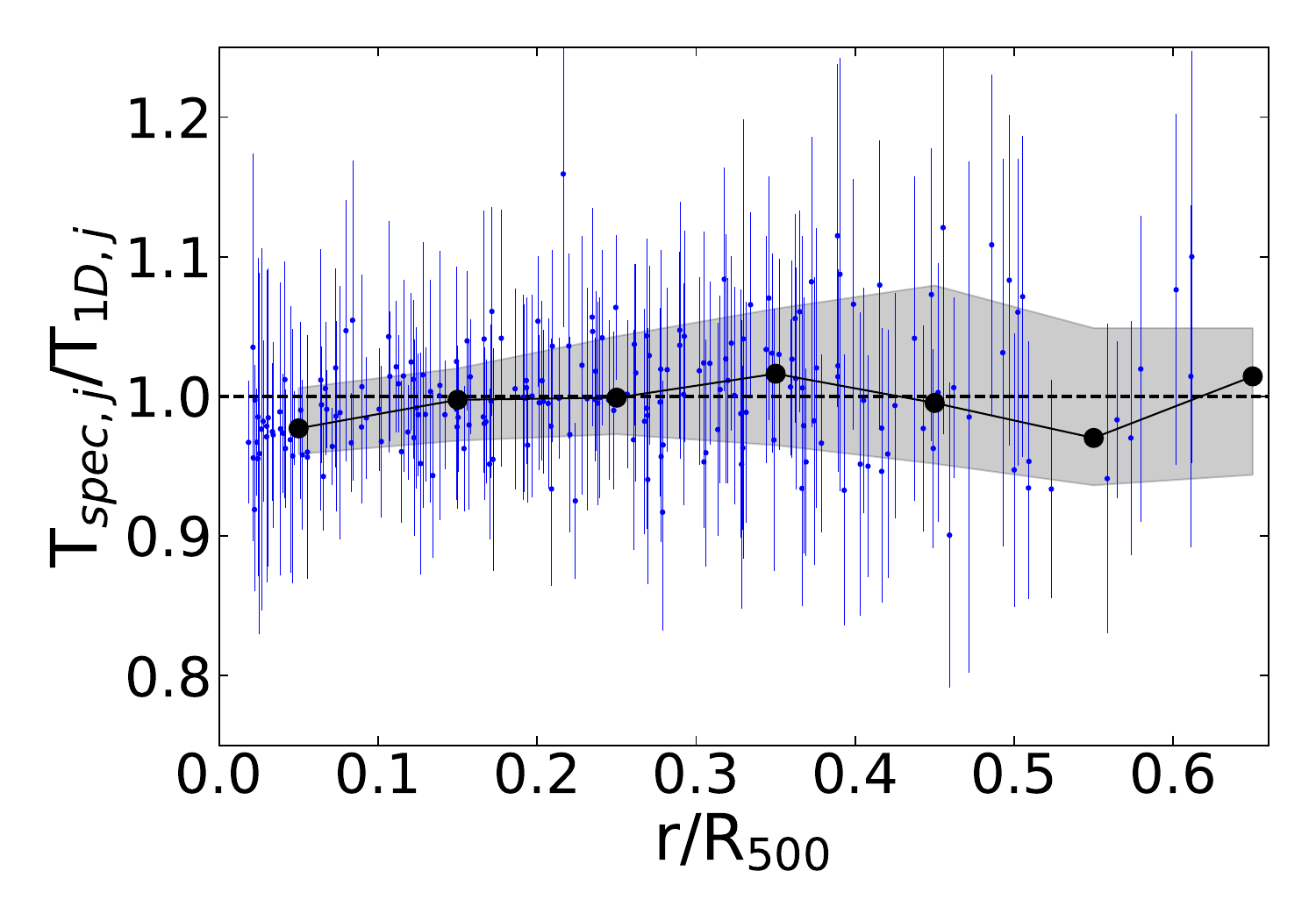}
\caption{Comparison between the profile derived 
 with the direct spectral fitting of each annulus $j$ (i.e., $T_{spec,j}$). and the temperatures recovered from the maps using Eq. \ref{T1D} (i.e., $T_{1D,j}$) in the same spatial regions. Each blue point corresponds to an annulus of the 28 clusters, while the black points represents the average values with the shadow area including the 16th and 84th percentiles. }
\label{fig:T1DTprof}
\end{figure}

We also associate to each measurement of $T_{1D, j}$ an uncertainty $\epsilon_{T_{1D,j}}$ from the propagation of the error $\epsilon_{T_{2D,i}}$ (we assume symmetric errors) on the spectral measurement $T_{2D, i}$
\begin{equation} \label{eq:epsilon}
\epsilon_{T_{1D,j}} = \frac{\sqrt{ \sum{ w_i^2 \epsilon_{T_{2D,i}}^2 \left[ (1+\alpha)\sum w_i-(\alpha/T_{2D, i}) \sum T_{2D, i} w_i \right]^2} }}{\left( \sum w_i \right)^2},
\end{equation}
and a standard deviation $\sigma_{T_j}$ (i.e., the total scatter or dispersion around the average value) defined as
\begin{equation} \label{eq:fluc_sigma}
\sigma_{T_j} = \sqrt{ \frac{\sum{ w_i \left(T_{2D, i} -T_{1D, j}\right)^2 }}{\sum{w_i}}}.
\end{equation}
The measured temperature dispersion in a given annulus is a combination of the intrinsic variation ($\sigma_{T_{j,int}}$) of the ICM temperature distribution (due to turbulent motions and mergers), of the statistical uncertainty $\epsilon_{T_{1D,j,stat}}$ on $T_{1D, j}$ (see Eq.~\ref{eq:epsilon}), and of the statistical uncertainty $\epsilon_{T_{i,stat}}$ of all the cells in the region of interest (as value we used the weighted mean of the statistical errors in a given region). 
Thus, the intrinsic temperature fluctuations are: 
\begin{equation} \label{eq:fluc_intsigma}
\sigma_{T_{j,int}} = \sqrt{\sigma_{T_j}^2 - \epsilon_{T_{1D,j,stat}}^2 - \epsilon_{T_{i,stat}}^2}.
\end{equation}

\begin{figure}[t]
\centering
\vbox{
\includegraphics[width=0.32\textwidth]{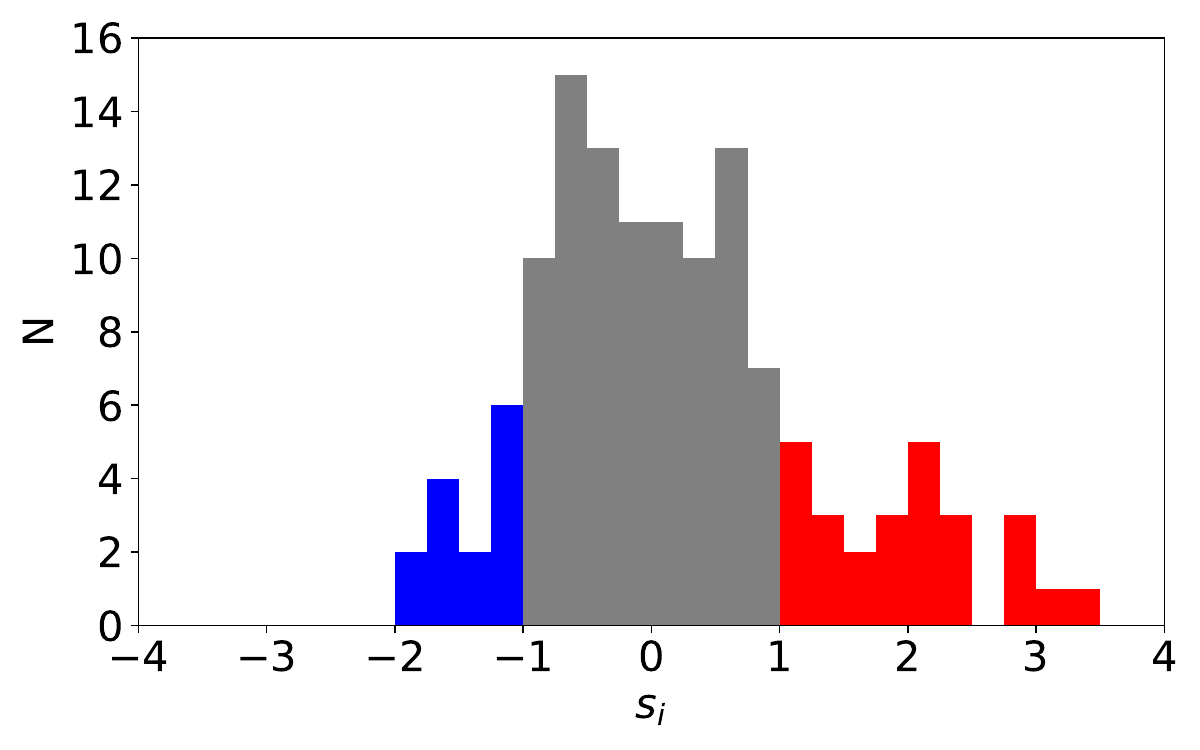}
\includegraphics[width=0.35\textwidth]{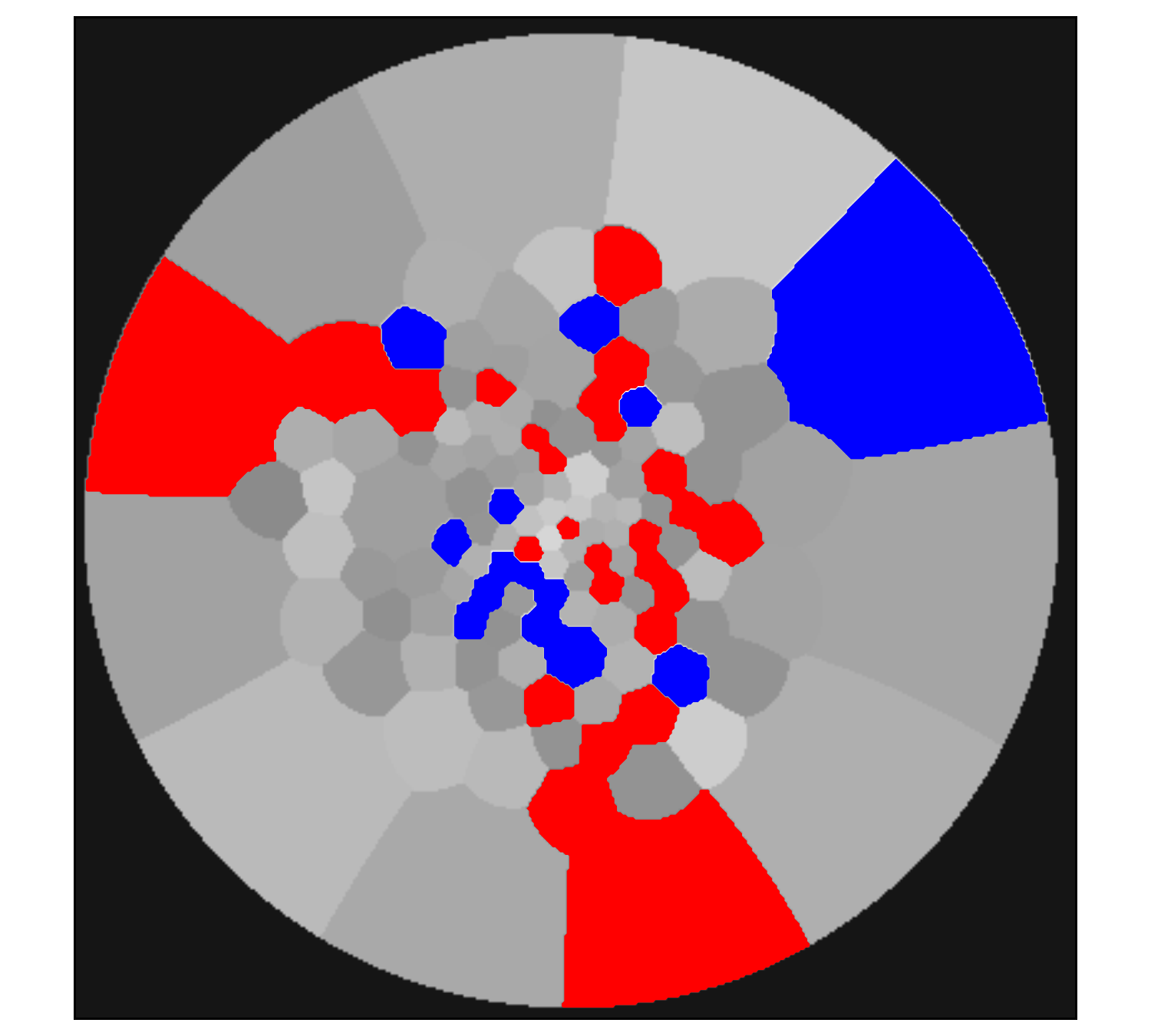}
}
\caption{Example of clipping for the cluster G041.45+29.10. In the {\it top} panel we show the distribution of $s_i$ values with the regions in red and blue being the cells deviating more than $1\sigma$ from the azimuthal value. In the {\it bottom} panel, we show their distribution in the temperature map derived within $R_{500}$.}
\label{fig:clipping}
\end{figure}

The gas density radial profile is recovered from the geometrical deprojection of the normalization of the spectral model (see Eq.~\ref{eq:n_k}).
Since the length along the line of sight is the same at a given radius (under the assumption of spherical symmetry), we can relate the line of sight averaged electron density at the annulus $j$ to the ones measured in each cell $i$ as 
\begin{equation}\label{eq:n_i_j}
n_j^2 A_j = \sum{n_i^2 A_{ij}}.
\end{equation}
From Eq.~\ref{eq:n_i_j}, we can then write
\begin{eqnarray}\label{ne1D}
n_{e,j} = & \sqrt{ \frac{\sum{n_{e,i}^2 A_{ij}}}{\sum{A_{ij}}} }, \nonumber \\
 \epsilon_{n_{e,j}} = & \frac{\sqrt{\sum{n_{e,i}^2 A_{ij}^2 \epsilon_{n_{e,i}}^2  }}}{n_{e,j} \sum{A_{ij}}}, \nonumber \\
\sigma_{n_{ej}}^2 = & \frac{\sum{ A_{ij} \left(n_{e,i} -n_{e,j}\right)^2 }}{\sum{A_{ij}}}.
\end{eqnarray}

We repeat the process after excluding the cells where the temperature deviates by more than 1$\sigma$  from the corresponding azimuthally averaged value $T_{{\rm 1D},j}$. 
This choice allow to remove also some of the noise in the maps which is related to the S/N criteria used to produce the maps and can be changed or improved with deeper observations.   
This $\sigma$--clipping is applied to the quantity
\begin{equation}
s_i = \frac{T_{{\rm 1D},j} - T_{2D,i}  }{\left( \epsilon_{T_{{\rm 1D}, j}}^2 + \epsilon_{T_{2D,i}}^2 \right)^{1/2}},
\label{eq:sigma-clip}
\end{equation}
with $T_{1D,j}$ (computed using Eq. \ref{T1D}) being the temperature in the annulus encompassing the cell of interest.
An example of such clipping is shown in Fig. \ref{fig:clipping}. If these regions are associated to cold clumps, turbulence, and/or bulk motions as suggested by simulations, then the recovered profiles will be more representative of a component in a nearly hydrostatic equilibrium. 
We discuss in Appendix \ref{multimaps} the impact of the Voronoi binning in detecting the temperature inhomogeneities.

\section{Results on the temperature maps}
\label{sect:res}
In this section, we present the results of the analysis done by estimating both the global values (i.e., within $R_{500}$) of the mean temperature and of the associated scatter, and of the same quantities resolved radially.
We study how these quantities distribute, how they relate among them, and if they can be used as a proxy of the dynamical state.
The relations between scatter and central values of the gas temperature (and density) are then used to investigate the physical properties of the local variations.

\begin{figure}[t]
\centering
\includegraphics[width=0.45\textwidth]{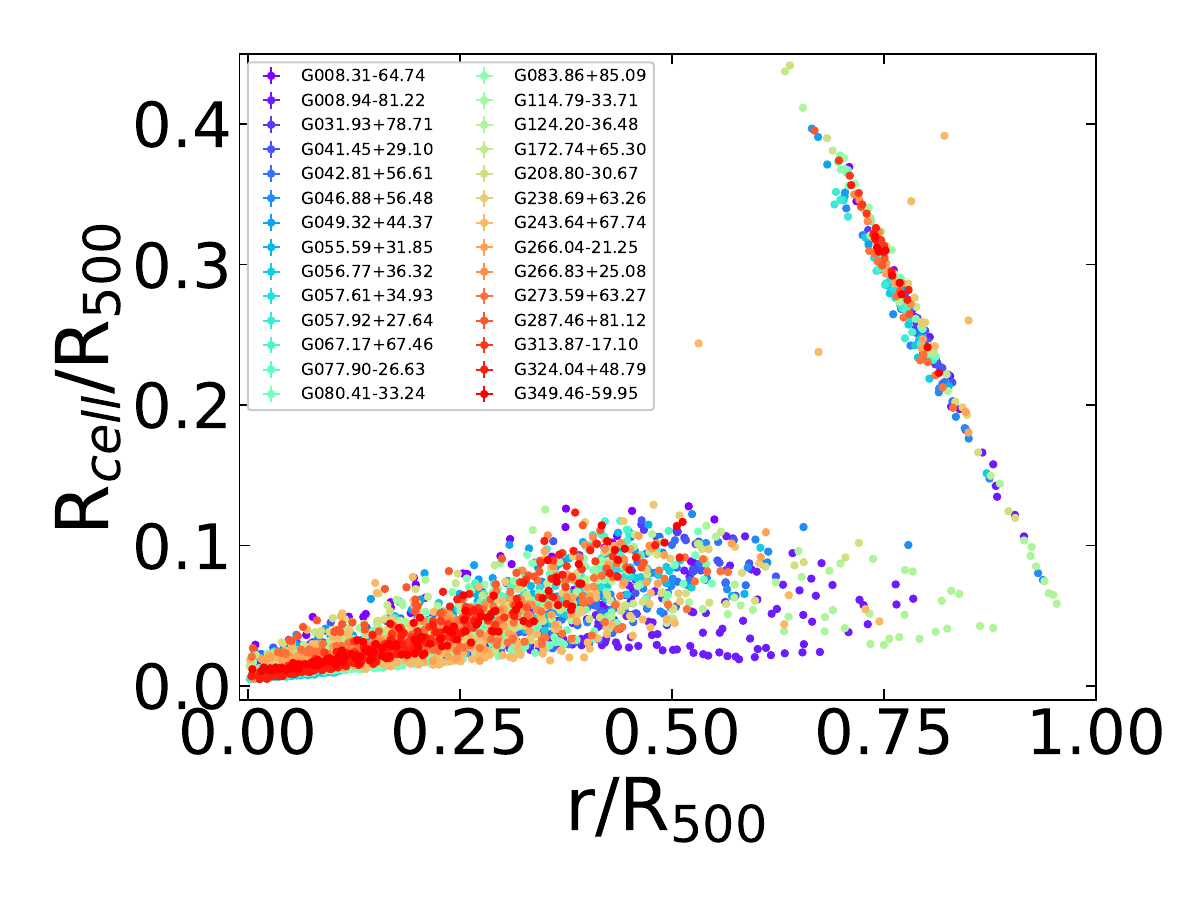}
\caption{
Radius of the cells for the maps obtained with S/N=30, in fraction of $R_{500}$, as a function of the distance from the center. The feature observed at $r/R_{500}\gtrapprox$0.65 is the result of the large cell sizes in the outer regions.}
\label{fig:binsize}
\end{figure}

\begin{figure*}[t]
\centering
\includegraphics[width=1\textwidth]{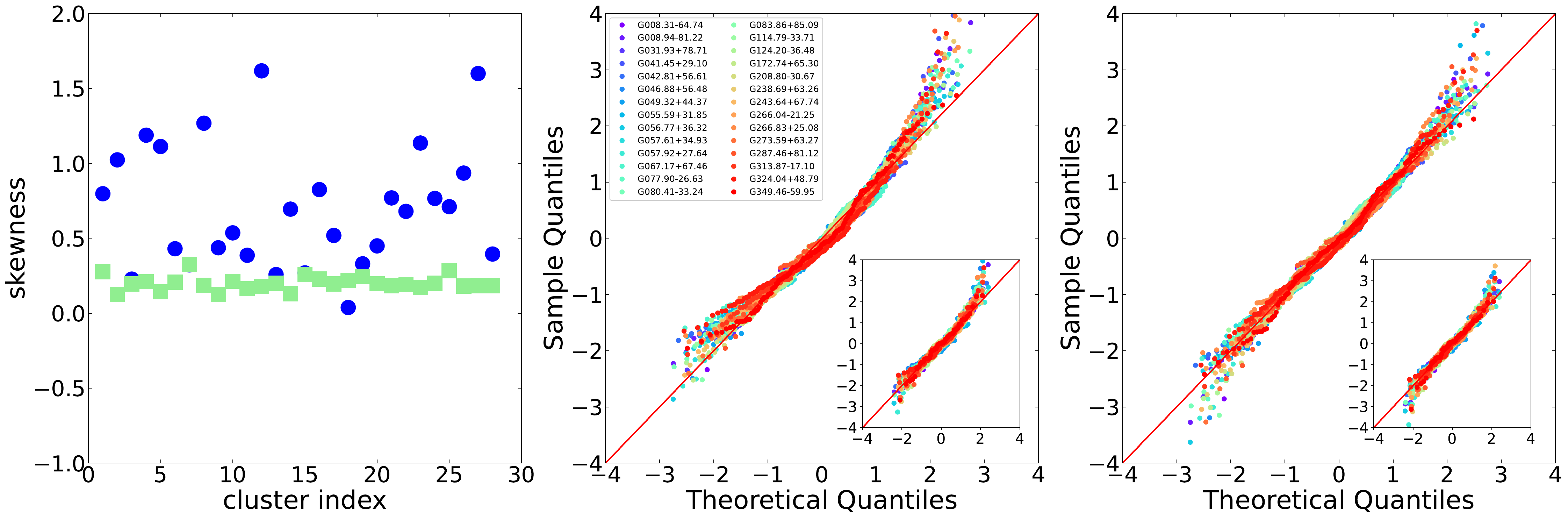}
\caption{
Observed properties of the 2D temperature distribution.
{\it Left:} measured values (blue circles) of the skewness in the temperature distribution of $T_{2D,i}/T_{1D,j}$ (being $T_{1D,j}$ the temperature of the annulus encompassing the cell $i$) compared with a typical uncertainty (green squares) that can be associated with the number of cells on the temperature map. {\it Middle:} QQ plot, assuming a normal distribution, of the $T_{2D,i}/T_{1D,j}$ values for the maps with S/N=30. {\it Right:}  QQ plot assuming a log-normal distribution. In the inset plots, we show the results when using the maps with S/N=50. The units of the QQ plots are given in z-score (i.e., subtracting the mean from each data point and dividing it by the standard deviation). }
\label{fig:skewQQ}
\end{figure*}

\begin{table*}[t!]
\centering
\caption{Parameters determined from the temperature maps with S/N=30. }
\label{tab:specpar}
\begin{tabular}{c c c c c c c c c c c c c }
\hline
\hline
\noalign{\smallskip}
Name & N$_{cells}$ & $T_{1D,500}$ & $\overline{|s_i|}$ & std($s_i$) & $\overline{\sigma_{T_{i, int}}}$ & $\overline{\sigma_{T_{i, int}}/T_j}$ & N$_{c}$ & $T_{1D,500,c}$ &  $\overline{|s_{i,c}|}$ & std($s_{i,c}$) & $\overline{\sigma_{T_{i,int,c}}}$ & $\overline{\sigma_{T_{i,int,c}}/T_{j,c}}$ \\
\noalign{\smallskip}
\hline
\noalign{\smallskip}
G008.31-64.74 & 59 & 6.75 & 0.65 & 0.84 & 1.12 & 0.16 & 47 & 6.83 & 0.42 & 0.50 & 0.73 & 0.10 \\ 
G008.94-81.22 & 333 & 8.62 & 1.03 & 1.46 & 2.20 & 0.25 & 219 & 9.04 & 0.48 & 0.57 & 1.23 & 0.13 \\ 
G031.93+78.71 & 157 & 3.07 & 1.21 & 1.56 & 0.55 & 0.16 & 84 & 3.17 & 0.48 & 0.56 & 0.27 & 0.08 \\ 
G041.45+29.10 & 130 & 6.25 & 0.87 & 1.12 & 1.22 & 0.19 & 90 & 6.41 & 0.45 & 0.53 & 0.60 & 0.09 \\ 
G042.81+56.61 & 250 & 4.67 & 1.04 & 1.34 & 1.01 & 0.20 & 154 & 4.67 & 0.46 & 0.54 & 0.45 & 0.09 \\ 
G046.88+56.48 & 119 & 5.12 & 0.90 & 1.16 & 1.00 & 0.19 & 81 & 5.17 & 0.50 & 0.58 & 0.54 & 0.10 \\ 
G049.32+44.37 & 44 & 4.94 & 0.85 & 1.05 & 0.83 & 0.15 & 30 & 5.00 & 0.48 & 0.57 & 0.50 & 0.09 \\ 
G055.59+31.85 & 159 & 7.45 & 0.75 & 0.97 & 1.23 & 0.16 & 112 & 7.90 & 0.42 & 0.50 & 0.80 & 0.10 \\ 
G056.77+36.32 & 334 & 5.09 & 0.86 & 1.12 & 0.77 & 0.15 & 220 & 5.19 & 0.44 & 0.53 & 0.41 & 0.08 \\ 
G057.61+34.93 & 105 & 4.38 & 0.99 & 1.22 & 0.88 & 0.18 & 59 & 4.63 & 0.42 & 0.51 & 0.42 & 0.09 \\ 
G057.92+27.64 & 198 & 3.29 & 1.12 & 1.64 & 0.49 & 0.14 & 117 & 3.46 & 0.43 & 0.53 & 0.25 & 0.07 \\ 
G067.17+67.46 & 176 & 8.66 & 0.93 & 1.22 & 2.31 & 0.25 & 113 & 9.02 & 0.44 & 0.52 & 1.15 & 0.12 \\ 
G077.90-26.63 & 130 & 5.17 & 0.90 & 1.15 & 0.80 & 0.15 & 88 & 5.38 & 0.47 & 0.55 & 0.46 & 0.08 \\ 
G080.41-33.24 & 318 & 5.89 & 0.95 & 1.24 & 1.50 & 0.23 & 203 & 6.20 & 0.48 & 0.57 & 0.75 & 0.11 \\ 
G083.86+85.09 & 74 & 5.56 & 0.89 & 1.22 & 1.05 & 0.18 & 48 & 5.60 & 0.38 & 0.48 & 0.50 & 0.08 \\ 
G114.79-33.71 & 96 & 5.22 & 0.78 & 0.98 & 0.89 & 0.16 & 66 & 5.48 & 0.42 & 0.50 & 0.48 & 0.08 \\ 
G124.20-36.48 & 162 & 5.46 & 1.19 & 1.53 & 1.23 & 0.22 & 78 & 5.71 & 0.40 & 0.49 & 0.52 & 0.09 \\ 
G172.74+65.30 & 115 & 3.98 & 0.93 & 1.17 & 0.69 & 0.16 & 72 & 4.06 & 0.47 & 0.55 & 0.38 & 0.09 \\ 
G208.80-30.67 & 84 & 6.53 & 0.93 & 1.31 & 1.27 & 0.19 & 52 & 6.79 & 0.38 & 0.46 & 0.59 & 0.09 \\ 
G238.69+63.26 & 135 & 4.79 & 1.05 & 1.42 & 0.92 & 0.18 & 75 & 4.97 & 0.39 & 0.47 & 0.37 & 0.07 \\ 
G243.64+67.74 & 179 & 4.88 & 0.89 & 1.13 & 0.83 & 0.17 & 119 & 5.00 & 0.50 & 0.57 & 0.60 & 0.12 \\ 
G266.04-21.25 & 142 & 10.71 & 0.90 & 1.24 & 2.65 & 0.24 & 95 & 11.34 & 0.43 & 0.51 & 1.43 & 0.12 \\ 
G266.83+25.08 & 169 & 6.21 & 0.73 & 0.90 & 0.89 & 0.14 & 127 & 6.10 & 0.47 & 0.54 & 0.51 & 0.08 \\ 
G273.59+63.27 & 144 & 5.55 & 1.07 & 1.56 & 1.25 & 0.21 & 93 & 5.88 & 0.52 & 0.58 & 0.73 & 0.11 \\ 
G287.46+81.12 & 54 & 2.85 & 0.88 & 1.26 & 0.58 & 0.17 & 34 & 3.31 & 0.37 & 0.44 & 0.33 & 0.09 \\ 
G313.87-17.10 & 153 & 8.93 & 0.78 & 1.04 & 1.89 & 0.20 & 113 & 9.25 & 0.47 & 0.56 & 1.20 & 0.12 \\ 
G324.04+48.79 & 185 & 11.60 & 0.82 & 1.05 & 3.11 & 0.25 & 129 & 11.60 & 0.46 & 0.54 & 1.76 & 0.15 \\ 
G349.46-59.95 & 155 & 11.58 & 0.80 & 1.08 & 2.50 & 0.21 & 113 & 12.32 & 0.45 & 0.53 & 1.60 & 0.13 \\ 
\noalign{\smallskip}
\hline
\end{tabular}
\tablefoot{
Cluster names are shown in column 1. In column 2, we provide the number of temperature cells for each cluster. In column 3, the overall temperature in keV recovered using Eq. \ref{T1D} in the annulus 0-$R_{500}$.
In columns 4–7, we provide the estimated spectroscopic parameters. Column 8 indicates the number of cells left after removing the ones deviating more than 1$\sigma$.
The corresponding temperature and  spectroscopic parameters are quoted in columns 9–13.
}
\end{table*}

\subsection{Properties of the 2D distribution}

The number of regions in the maps determined with S/N=30 varies from $\sim$50 to $\sim$300 per cluster (median $\sim$150; see Table \ref{tab:specpar}) depending on the quality of the data and the brightness of the cluster. 
The size of the cells as a function of the distance from the X-ray center (in units of $R_{500}$) is shown in Fig. \ref{fig:binsize}.
The distribution of the temperature and electron density values can provide some basic information about the physics of the ICM,  like the constraining of the turbulent velocity in galaxy clusters. 
Several numerical studies suggest that the temperature and electron density distributions are approximately log-normal (e.g., \citealt{Kawahara2007,zhu13,Frank2013,Rasia2014,gas14,Towler2023}). We tested whether the observed temperature distributions for our clusters differ from a normal distribution. 
For simplicity, we neglect the variations between the different rings and we fit the distribution for the whole cluster within $R_{500}$. In Fig \ref{fig:skewQQ} ({\it left} panel), we show the measured skewness of the distribution of $T_{2D,i}/T_{1D,j}$ (where $T_{1D,j}$ is the temperature in the  annulus encompassing the cell $i$) that should be close to zero for normal distributions. 
For all the clusters, we measure a positive skewness indicating a tail of higher temperature values that could be associated with heating events (e.g., shocks) that did not have time to thermalize. 
However, the chosen sampling of the map (i.e., the number of cells) can impact the measured values (i.e., the skewness measured in the presence of a low number of cells can be quite uncertain). 
To test the uncertainty associated with the number of cells present in the map for each cluster, we simulate 1000 distributions of N values (where N is equal to the number of cells in the map) randomly selected from a normal distribution. We then measure the skewness for each distribution and evaluate the dispersion around zero (i.e., around the expected value). 
The result is shown with green points in the {\it left panel} of Fig.~\ref{fig:skewQQ}, where typically the measured values are significantly larger than the uncertainty that can be associated with the limited numbers of cells available. 
That suggests that indeed the distributions have a tail of high-temperature values. 
The skewness values of the observed distribution do not change significantly even if we remove the core regions (i.e., $<0.15R_{500}$), typically cooler, that potentially drive the distribution. 
We note that, since the temperatures $T_{1D,j}$ are recovered from the measured $T_{2D,i}$ values (and not from a direct spectral fitting of the region), potential biased measurements in the latter quantity will affect as well the former (i.e., it is not expected to impact significantly the skewness). 

\begin{table}[t!]
\centering
\caption{Spearman correlation coefficient $r$ and $p-value$ between the spectral and morphological parameters.}
\label{tab:pars}
\setlength\tabcolsep{1.5pt}
\begin{tabular}{c | cccc}
\hline
\hline
\multicolumn{1}{c}{} & \multicolumn{4}{|c}{region}\\
\hline
Rel. & $<R_{500}$ & $<0.15R_{500}$  & $0.15-0.5R_{500}$ & $>0.5R_{500}$   \\
\hline
 & r (p-val) & r (p-val) & r (p-val) & r (p-val)  \\
\hline
$\overline{|s_i|}$--$c$ & -0.17(0.39) & 0.34(0.08) & -0.30(0.12) & -0.02(0.91) \\
$\overline{|s_i|}$--$w$ & 0.33(0.09) & 0.00(0.99) & 0.42(0.03) & -0.06(0.75) \\
$\overline{|s_i|}$--$P_{20}$ & 0.15(0.44) & -0.22(0.26) & 0.16(0.42) & 0.10(0.61) \\
$\overline{|s_i|}$--$M_{all}$ & 0.30(0.13) & -0.20(0.31) & 0.37(0.06) & 0.08(0.70) \\
\smallskip\\
std($s_i$)--$c$ & -0.23(0.24) & 0.41(0.03) & -0.27(0.16) & -0.02(0.93) \\
std($s_i$)--$w$ & 0.37(0.05) & -0.09(0.64) & 0.38(0.05) & -0.00(0.99) \\
std($s_i$)--$P_{20}$ & 0.19(0.33) & -0.27(0.17) & 0.10(0.61) & 0.12(0.55) \\
std($s_i$)--$M_{all}$ & 0.35(0.07) & -0.26(0.17) & 0.32(0.09) & 0.11(0.58) \\
\smallskip\\
$\overline{\sigma_{T_{i,int}}/T_j}$--$c$ & -0.11(0.56) & 0.04(0.82) & -0.04(0.86) & 0.22(0.27) \\
$\overline{\sigma_{T_{i,int}}/T_j}$--$w$ & 0.16(0.40) & -0.02(0.93) & 0.16(0.41) & -0.20(0.30) \\
$\overline{\sigma_{T_{i,int}}/T_j}$--$P_{20}$ & -0.05(0.79) & -0.14(0.47) & -0.15(0.45) & -0.13(0.51) \\
$\overline{\sigma_{T_{i,int}}/T_j}$--$M_{all}$ & 0.15(0.46) & -0.00(0.98) & 0.05(0.80) & -0.14(0.47) \\
\hline
\end{tabular}
\tablefoot{
The coefficient $r$ is the non-parametric statistical measure used to study how well the relationship between two variables can be described using a monotonic function. 
The p-value for a hypothesis test whose null hypothesis is that two sets of data are linearly uncorrelated. 
}
\end{table}

A way to graphically show the deviation from a normal distribution is through the quantile-quantile (QQ) plots, where a quantile is the fraction of points below a given value. 
If two sets  come from a population with the same distribution, then the points should fall approximately along the 45-degree reference line. 
The greater the departure from this reference line, the greater the evidence that the two data sets have come from populations with different distributions. 
In the middle panel of Fig. \ref{fig:skewQQ}, we compare the temperature distribution of the $T_{2D,i}/T_{1D,j}$ values of each cluster with the normal distribution, and we see that the QQ plot for most of the clusters appears curved above the line. 
This shape indicates a distribution that is heavily right-skewed (i.e., hot clumps) with a light left tail (only partially associated with the cooler cells in the core region because some are associated with substructures that have merged). 
When comparing the logarithm of the temperatures with the lognormal distribution, we see that the QQ plots appear as roughly a straight line, suggesting that indeed the projected temperatures are roughly log-normally distributed. 
The Anderson-Darling test (more sensitive to the tails of the distribution with respect to the Kolmogorov-Smirnov test which is instead more sensitive to the center of distribution) indicates that the hypothesis of the normal distribution can be rejected at a significance level of 5\% (1\%) for 20 (14) clusters while the log-normal distribution only for 5 (4) clusters. 
We have also verified that similar results are obtained by excluding the innermost regions (i.e., the effect is not driven by the core where we have most of the map cells) or the outermost regions (where the cluster temperatures decline). 
By clipping out the regions with $|s_i|>1$ the distribution becomes closer to the normal distribution, especially, if the core regions are not considered. Similar results, although less significant, are obtained with coarser but higher S/N maps (see the inset plots of Fig. \ref{fig:skewQQ}).

\subsection{Temperature dispersion and dynamical state}

Radiative cooling, AGN feedback, and mergers leave clear imprints on the thermodynamic properties (e.g., temperature) of the ICM at different scales. 
However, the two-dimensional structures observed in the temperature maps have been mainly used so far for qualitative analysis (e.g., the determination of a shock or a cold front position) leaving their full potential unexploited. 

To investigate possible connections between the temperature dispersion and the cluster dynamical state,  we use the morphological parameters determined from the imaging analysis \citep[see][]{cam22}. 
In particular, we consider the morphological parameters $c$ (i.e., the ratio of the emission within two different apertures; more sensitive to the core properties), $w$ (i.e., the variance of the projected separation between the X-ray peak and the centroid of the emission obtained within 10 apertures; more sensitive to substructures), $P_{20}$ (i.e., a measure of the ellipticity based on the second moment of the power ratios which consist of a 2D multipole decomposition of the surface brightness distribution), and their combination $M_{\rm all}$. 
We refer to \cite{cam22} for the definition of these parameters.

For each temperature map cell, we computed $s_i$  given in Eq.~\ref{eq:sigma-clip}.
We then computed for each cluster an average $\overline{|s_i|}$ value and the standard deviation of $s_i$. The average values are given in Table \ref{tab:specpar}. 
In Table \ref{tab:pars}, we show the correlation of these values, in different cluster regions, with morphological parameters coming from the analysis of the X-ray images tabulated in Table \ref{tab:clusters} (i.e., computed within $R_{500}$). 
Apart from a moderate correlation of the spectroscopic parameters with $w$ (more related to the gas inhomogeneities than $c$, which is more related to the presence of strong cores) and $M_{\rm all}$ in the intermediate regions of the clusters (i.e., 0.15-0.5R$_{500}$), there is no (or very weak) correlation between the parameters from the spectral and imaging analyses. 
Although there is a mild anti-correlation (i.e., $r\sim$0.4) between the spectroscopic derived parameters and the cluster temperature that could potentially hide a moderate correlation between $s_i$ and the morphological parameters, it seems that the impact of the dynamical state on the temperature dispersion is subdominant with respect to other effects. 
The lack of correlation (or the difficulty to detect it) is probably related to the small range of radial temperature values ($\sim$90\% of the $T_{2D,i}$ cell values lie in the range 0.5-1.5$T_{500}$), or to the size of the cells that may be sensitive only to certain fluctuations (e.g., large scale). 
This is different from the gas density which varies by $\sim$3 orders of magnitude in the same radial range and is, therefore, more sensitive to local fluctuations induced from, for example, displacement of the gas, as occurs during mergers that are expected to be the primary source of the injection of extra-energy and turbulence. 
This reinforces the evidence that gas density-based proxies provide more powerful estimators of the dynamical state of massive halos (thanks also to the higher resolution that can be achieved).  

\begin{figure}[t]
\centering
\includegraphics[width=0.5\textwidth]{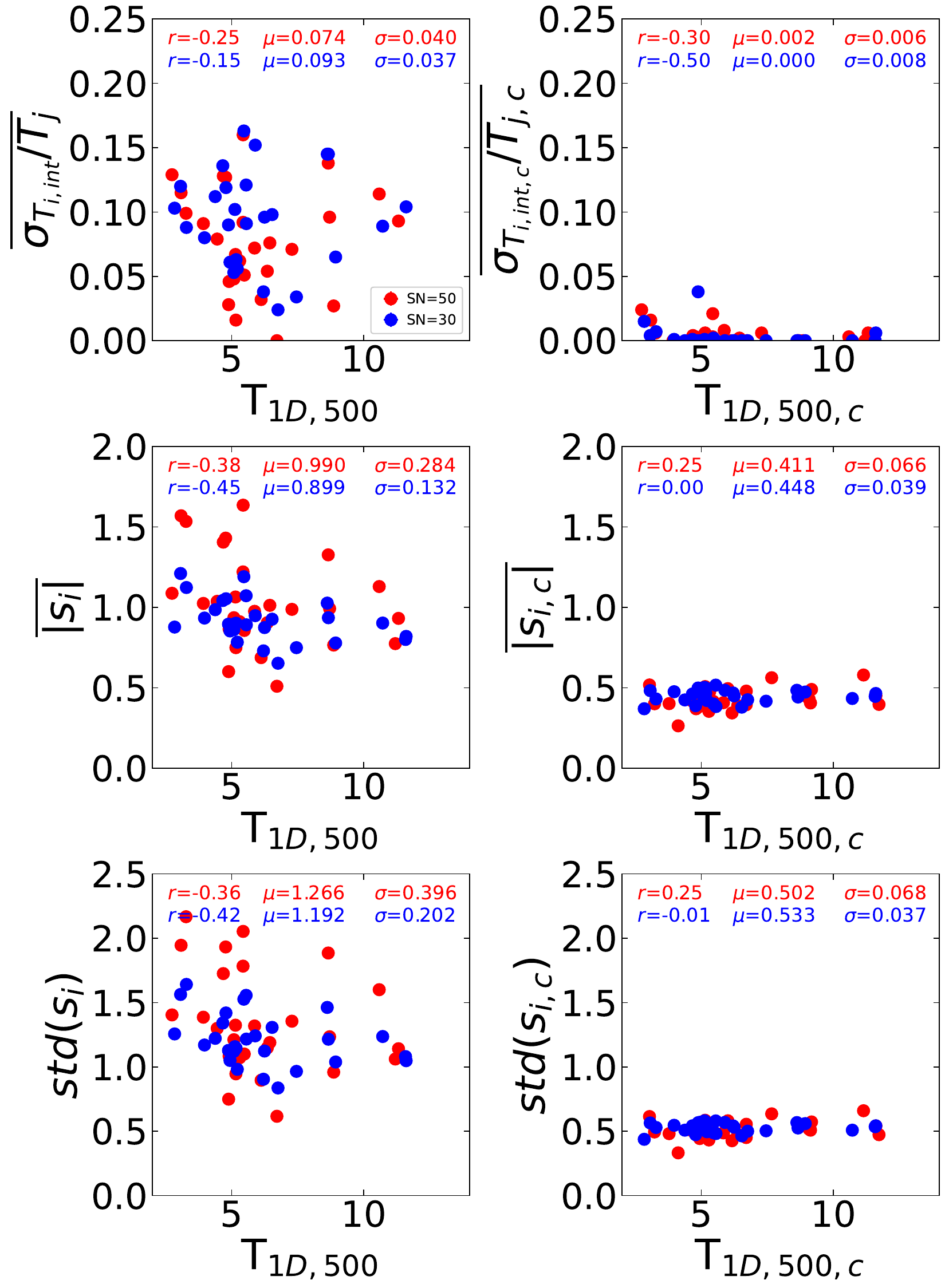}
\caption{Spectral parameters derived before ({\it left panels}) and after ({\it right panels}) clipping for the regions deviating more than $1\sigma$ from the azimuthal values. In blue and red we show the results obtained with maps of S/N=30 and S/N=50, respectively. In each panel we provide the correlation (i.e. Spearman rank test value $r$) between the spectral parameter and $T$, the median value of the distribution $\mu$, and the standard deviation $\sigma$.}
\label{fig:sn50sn30}
\end{figure}

For each cluster we also obtained the mean relative scatter $\overline{\sigma_{T_{i,int}}/T_j}$ of the temperature (see Table \ref{tab:specpar}) using Eq. \ref{eq:fluc_sigma} where the azimuthal value, $T_j$ at the radius of the cell is computed using Eq. \ref{T1D}. 
Again, there is no clear correlation with the morphological parameters.  

We repeated the calculation after removing all the cells with $|s_i|>1$ and we report the values in Table \ref{tab:specpar}. 
We note that, when considering the full maps, the parameters span a broad range of values, while after clipping there is a convergence toward constant values (see also Fig. \ref{fig:sn50sn30}). 
In particular, $\overline{\sigma_{T_{i,int}}/T_j}$ goes to zero as expected. 
The average values of $\overline{|s_i|}$ also decreases by a factor of $\sim$2 with a much smaller standard deviation and the removal of the moderate correlation with the cluster temperature.  
A similar effect is seen in the values of std($s_i$).  
It is important to note that the values decrease significantly in all clusters independently of their morphological appearance and that after clipping both $\overline{|s_i|}$ and std($s_i$) tend to converge to similar values independently of the resolution of the maps (i.e., S/N=30 vs S/N=50).

\begin{figure}[t]
\centering
\includegraphics[width=0.45\textwidth]{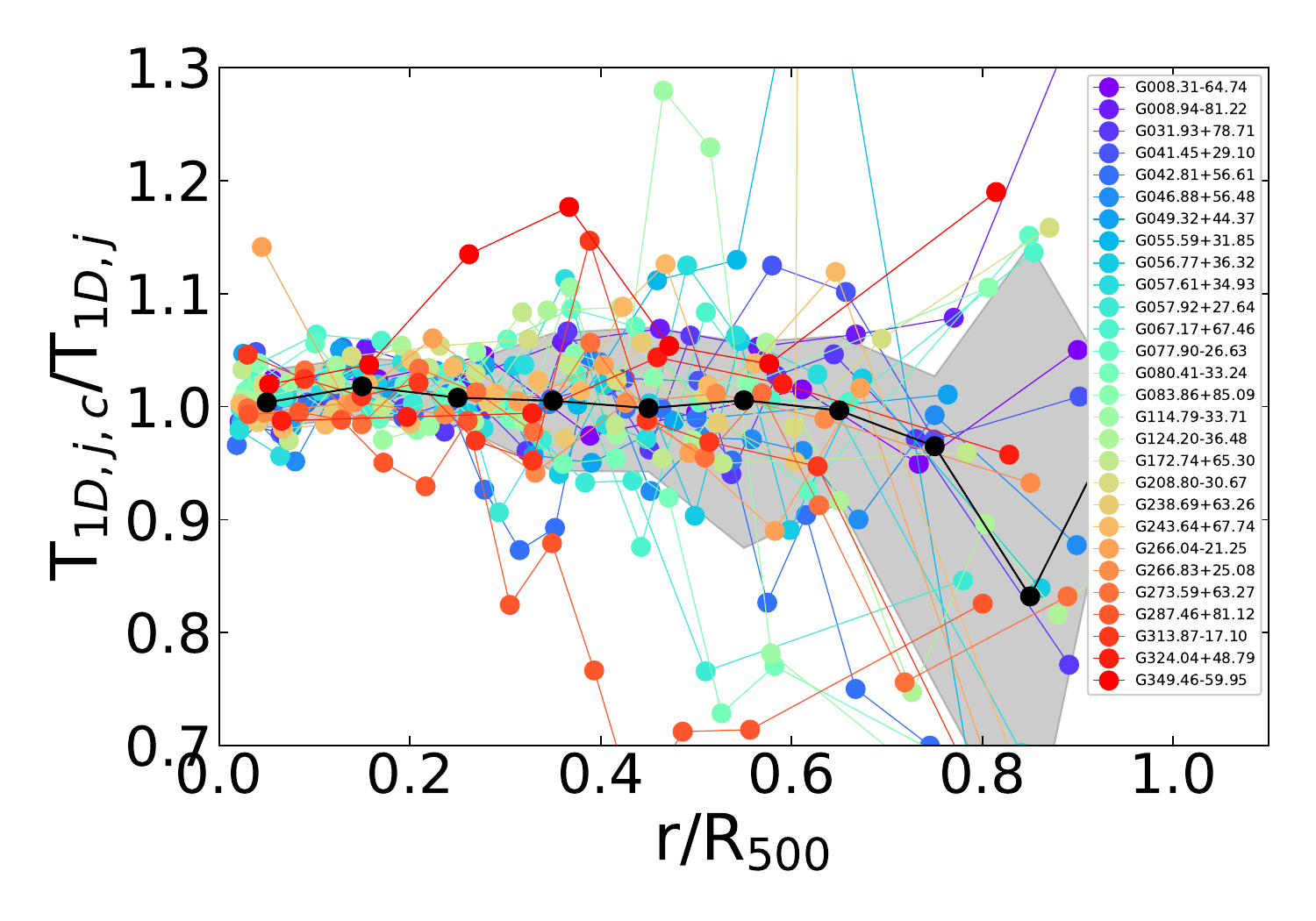}
\caption{Ratio between $T_{1D,j,c}$ (i.e., the temperature estimated in each annulus $j$ after clipping the most deviating regions) and $T_{1D,j}$ (i.e., the temperature estimated using all the temperature cells in each annulus) as a function of radius. The black points represents the average values, while the shadow area includes the 16th and 84th percentiles. 
}
\label{fig:Tclip}
\end{figure}

\subsection{Impact of the temperature fluctuations in the recovered azimuthal profile}

Removing the temperature inhomogeneities (associated with cold and hot clumps and substructures) should provide profiles that are closer to the expectation for gas in hydrostatic equilibrium with the gravitational potential.

The quantity $s$ (Eq.~\ref{eq:sigma-clip}) can be used to exclude regions that deviate from the azimuthal value. 
To highlight the impact, we excluded all the cells with $|s_i|>$1 before reconstructing the temperature profile. 
The fraction of cells that deviate more than 1$\sigma$ from the azimuthal value show no significant correlation with the dynamical state; the Spearman rank test between the fraction of scattered temperature cells and the $M_{\rm all}$ parameter  does not show evidence of a strong correlation ($r\sim$0.32; p-value=0.10). 
Even morphologically relaxed systems (i.e., high $c$ and low $w$ or $M_{all}$) do not have $T_{1D,j,c}/T_{1D,j}\sim1$.

In Fig. \ref{fig:Tclip}, we show the impact on the temperature measurements after removing the cells with $|s_i|>$1. 
In many central annuli, the new temperatures are slightly larger than the original value (a sign that we are slightly preferentially removing the, easier to detect, cold clumps). 
In general, within $\sim$0.3-0.4$R_{500}$, the effect is somehow small (i.e., $<$5\%) but at larger radii, the effect can be of 10-20\% or more in some particular regions.  

\begin{figure}[t]
\centering
\vbox{
\includegraphics[width=0.35\textwidth]{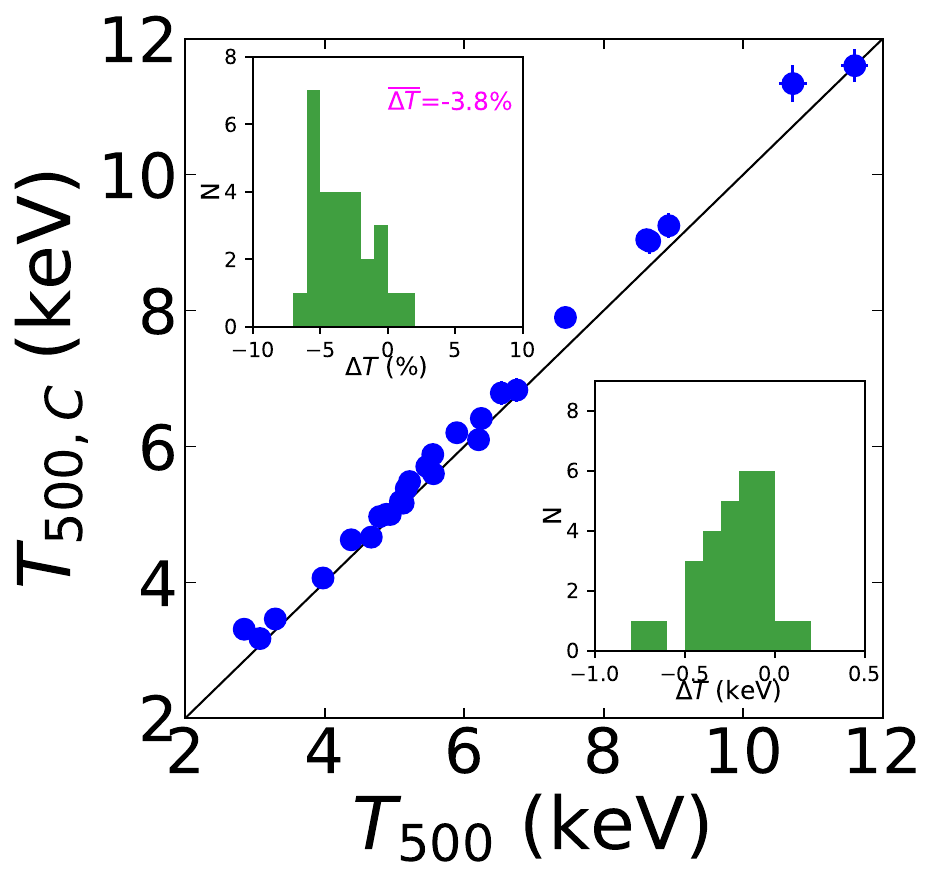}
\includegraphics[width=0.35\textwidth]{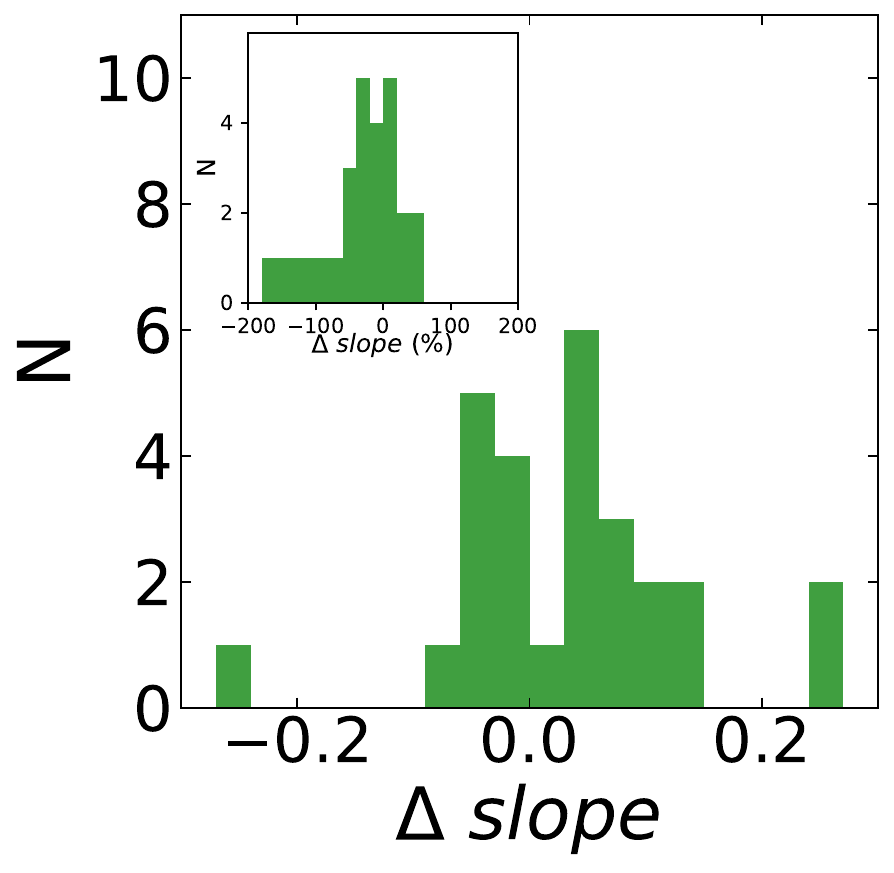}
}
\caption{
Impact of the temperature fluctuations to the cluster overall temperature and azimuthal profile.
{\it Top:} comparison between the global temperature estimated using all the cells in the 2D maps and the temperature estimated after excluding the cells deviating more than 1$\sigma$ with respect to the azimuthal value. In the inset plots we show the absolute difference between $T_{500}$ and $T_{500,c}$ ({\it bottom-right} inset) and its fractional variation ({\it top-left} inset). {\it Bottom:} change in the gradient of the temperature profile fitted in the region 0.15-0.75$R_{500}$ before and after the clipping. In the inset plot, we show its fractional variation.}
\label{fig:Tchange}
\end{figure}

This could have an obvious impact, for instance, on the mass reconstruction under the HE assumption and on the use of the overall cluster temperature in scaling relations (e.g., $M_{tot}$-$T$ or $M_{tot}$-$Y_X$). In fact, the recovered total mass depends linearly on the temperature at $R_{500}$ but also on its gradient. 
We are not in the position to quantify the real impact with our data because at large radii the binning of the maps is quite large and prevents a proper recovering of the inhomogeneities. 
However, in Fig. \ref{fig:Tchange} ({\it top panel}), we show how the global temperature changes after removing cells with $|s_i|>$1. 
There are several clusters showing a difference of $\sim$0.5 keV, corresponding to a relative change of $\sim$5\%. 
Assuming a self-similar $M_{tot}$-$T$ scaling, this relative change in temperature would  correspond to a change in the total mass of $\sim$7-8\%.  
In Fig. \ref{fig:Tchange} ({\it bottom panel}), we show the change in slope of the temperature profile when fitting a power-law in the region 0.15-0.75$R_{500}$ (beyond 0.75$R_{500}$ the map is too coarse). 
There is a very mild correlation (i.e., r=-0.24) between the change in slope and in temperature.

\begin{figure*}[!t]
\centering
\includegraphics[width=1\textwidth]{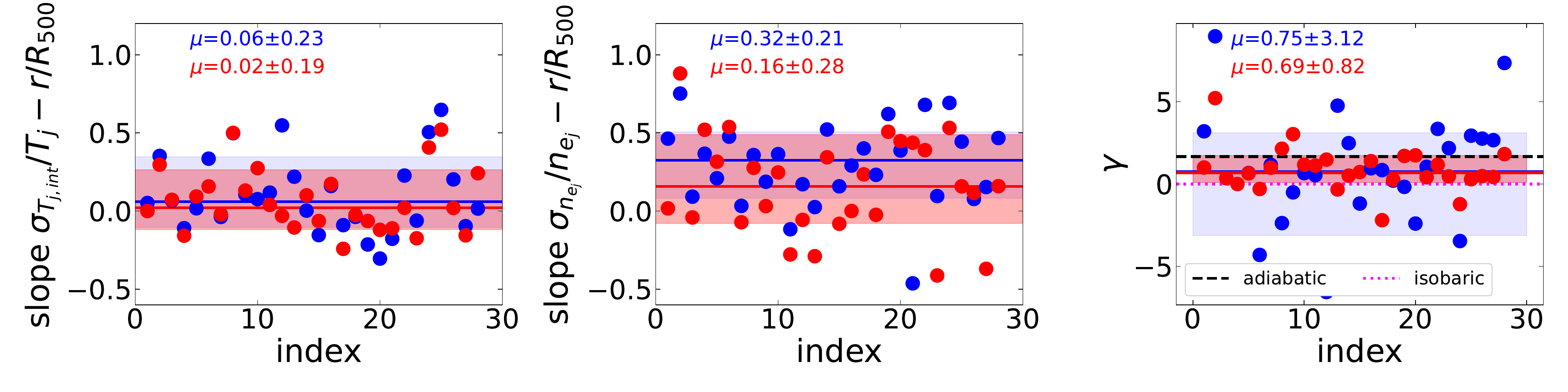}
\caption{
Slope values of the intrinsic scatter of the temperature ({\it left panel}) and electron density ({\it middle panel}) measured within $\sim$0.6$R_{500}$. In the {\it right} panel, we show the polytropic index $\gamma$ (see Eq. \ref{eq:fluc}) derived for each cluster. Blue points are from maps with S/N=30 and red points are from maps with S/N=50. The solid lines represent the median value $\mu$ (also reported in red on the top-left corners) while the shaded area includes the 16th and 84th percentiles.
}
\label{fig:scatterR}
\end{figure*}

\subsection{Properties of the fluctuations}\label{sec:type}

Since the pioneering work by \cite{sch04} that detected a scale-invariant pressure fluctuation spectrum in the range between 40 and 90 kpc in the Coma cluster well described by a projected Kolmogorov/Oboukhov-type turbulence spectrum, several works have tried to model the expected signal associated with the propagation of large-scale eddies due to the diffusion of energy injected from dynamical phenomena like mergers and mass accretion, and have put interesting constraints from the observed fluctuations in both the density distribution of the X-ray emitting gas \citep[e.g.,][]{chu12,are12,gas13,gas14,zhu14,zhu16,hof16} and in the SZ-based pressure maps \citep[e.g.,][]{kha16}.

In the present work, we complement previous studies with the analysis of the temperature fluctuations measured in the (projected) temperature maps, under the assumption that the projected fluctuations provide a proxy for the turbulence field (as done in, e.g., \citealt{sch04,hof16,kha16}). 
To support this assumption, we computed also the second-order structure function (SF) of the temperature. 
The SF (see a definition in, e.g., \citealt{zuhone2016}) is directly related to the power spectrum of the projected quantity of interest, and has the advantage that it can be computed from the 2D distributions independently from the spatial shape of the regions under consideration. 
We focused on small scales ($r<$150 kpc) to avoid the flattening due to the large-scale plateau (see, e.g., \citealt{Roncarelli2018}), and measured an average slope of the SF of $\sim$0.8--1. 
Although this is flatter than the slope of 5/3 expected for a Kolmogorov-like power spectrum, it is roughly in agreement with the finding of \citet{Roncarelli2018}, where the overall shape of the SFs derived in constrained hydrodynamical simulations are less steep than the expectations from the adopted power spectra. 
It is known that measuring temperature fluctuations is more complicated than measuring density fluctuations (from surface brightness) and a larger S/N is required. 
Clearly, this translates to a coarser temperature mapping, so that different physical scales are probed with respect to the SB maps. 
In this case, the spectroscopically estimated $T_{2D,i}$ is the result of the emission-weighted\footnote{The spectroscopic temperatures tend to highlight cooler regions, and thus the micro fluctuations. 
Conversely, the mass- and volume-weighted temperatures often used in simulations tend to make the projected maps more homogeneous, with the latter being able to diminish even the peaks in the cluster core (a few 100 kpc). 
This will likely translate into a larger scatter for the spectroscopic temperatures but will also help to trace the 2D features.} average along the line of sight of the three-dimensional temperature $T$ (see, e.g., Mazzotta et al. 2004). 
For a polytropic gas, fluctuations in the latter quantity, $\delta T_i = T_i-\overline{T}$, should satisfy the relation 
\begin{equation}
    \frac{\delta T}{T}= (\gamma -1) \frac{\delta n}{n},
    \label{eq:fluc}
\end{equation}
with the fluctuations in the gas density $n$, where $\gamma$ is the polytropic index and is equal 0 and 5/3 for the isobaric and adiabatic case, respectively.   Following the arguments in \cite{sch04}, we can convert these three-dimensional fluctuations into the corresponding projected quantities once the temperature is considered as a weighted quantity along the line of sight
\begin{equation}
    \frac{\delta T_{2D}}{T_{2D}} \approx 
    \frac{\int dl \, w \ \delta T}{\int dl \, w \, T}  \approx \frac{\gamma -1}{2} \frac{\delta n_{2D}^2}{n_{2D}^2},
    \label{eq:fluc2D}
\end{equation}
where $w \approx n^2 T^a$, $n_{2D}^2 = \int dl \, n^2$, and $\delta n_{2D}^2=2n_{2D} \delta n_{2D}$. Here, we are assuming that within each cell, a single temperature and density, with their associated fluctuations, are present.  It is worth noting that projection effects might be more complex than the assumption adopted here, implying the calculation of a proper window function in the band of interest for each object \cite[see, e.g.,][]{kha16}. 
However, we show in Appendix \ref{2D3Deff} that  the relation between Eq.~\ref{eq:fluc} and \ref{eq:fluc2D} works pretty well for a sample of simulated objects.

In our analysis, we identify as fluctuations in temperature the dispersion  (see Eq.~\ref{eq:fluc_sigma}) of the measured spectral temperature around the average cluster temperature in the region of interest. Therefore, we fitted the relation:
\begin{equation}
\frac{\sigma_{T_{j,int}}}{T_j}=\frac{\gamma -1}{2} \left( \frac{\sum{n_{e,i}^2 A_{i,j}}}{n_{e,j}^2} -1 \right).
\label{eq:fitbeta}
\end{equation}

In the {\it left panel} of Fig.~\ref{fig:scatterR}, we show the radial trend of the temperature scatter measured by fitting a power-law to the radial cells of each cluster with a size of 0.1$R_{500}$ up to a radius where the S/N=30 maps have enough resolution (typically 0.6$R_{500}$; see Fig. \ref{fig:binsize}). 
The temperature dispersion profiles are similar to one another and are in general quite flat (i.e., slope close to zero for most of the clusters). This result stands both for relaxed and disturbed clusters (i.e., independent of the dynamical state) and does not depend on the map resolution.

The fluctuations in density have been estimated similarly to the ones in temperature (see Eq. \ref{eq:fluc_sigma} and Eq. \ref{ne1D}) using the line of sight averaged electron density values estimated with Eq. \ref{eq:n_k}. 
The shape of the density distribution (e.g., the QQ plots) is similar to the one in temperature and to what is found in simulations (e.g., \citealt{zhu13}) with a heavy tail in the high-density regime (i.e., well-described by a log-normal distribution) usually associated with cold regions (e.g., substructures).
The absolute values of the scatter in density are quite large and are mainly driven by the poor resolution of the electron density maps (derived with the temperature map binning) paired with steep density gradients.  
In Appendix \ref{multimaps}, using surface brightness maps, we show how the fluctuations change as a function of the resolution. 
However, independent of the resolution, the scatter on the density for many clusters slightly increases with radius (see the slope values in the {\it middle panel} of Fig. \ref{fig:scatterR}) in agreement with the finding by simulations (see \citealt{Rasia2014,Towler2023}). 

By fitting Eq.~\ref{eq:fitbeta}, we can make a measure of $\gamma$ for each cluster. 
The recovered values are shown in the {\it right panel} of Fig.~\ref{fig:scatterR}. 
They do not depend on the map resolution (although the S/N=50 maps allow a better constraint thanks to their lower statistical uncertainties) or on the dynamical state (i.e., $r\sim$0.1 and $r\sim$0.2 with concentration and centroid-shift, respectively). 
They lie, on average, between the isobaric and the adiabatic expectations. The 1$\sigma$ interval of derived $\gamma$ is large and also overlaps with $\gamma$=1 (although in our sample there are only 4 clusters with $\gamma$ in the range [0.8-1.2]). 
However, the pure isothermal state is a very narrow, hard to achieve regime, since turbulence is continuously regenerated (see details in \citealt{gas14}).
Obviously, the perturbations in real clusters are a mix of the different regimes and the $\gamma$ values reflect such mixing. 
The values of $\gamma$ do not show any correlation with morphological parameters, total mass, or redshift.

\begin{figure*}[t]
\centering
\includegraphics[width=0.8\textwidth]{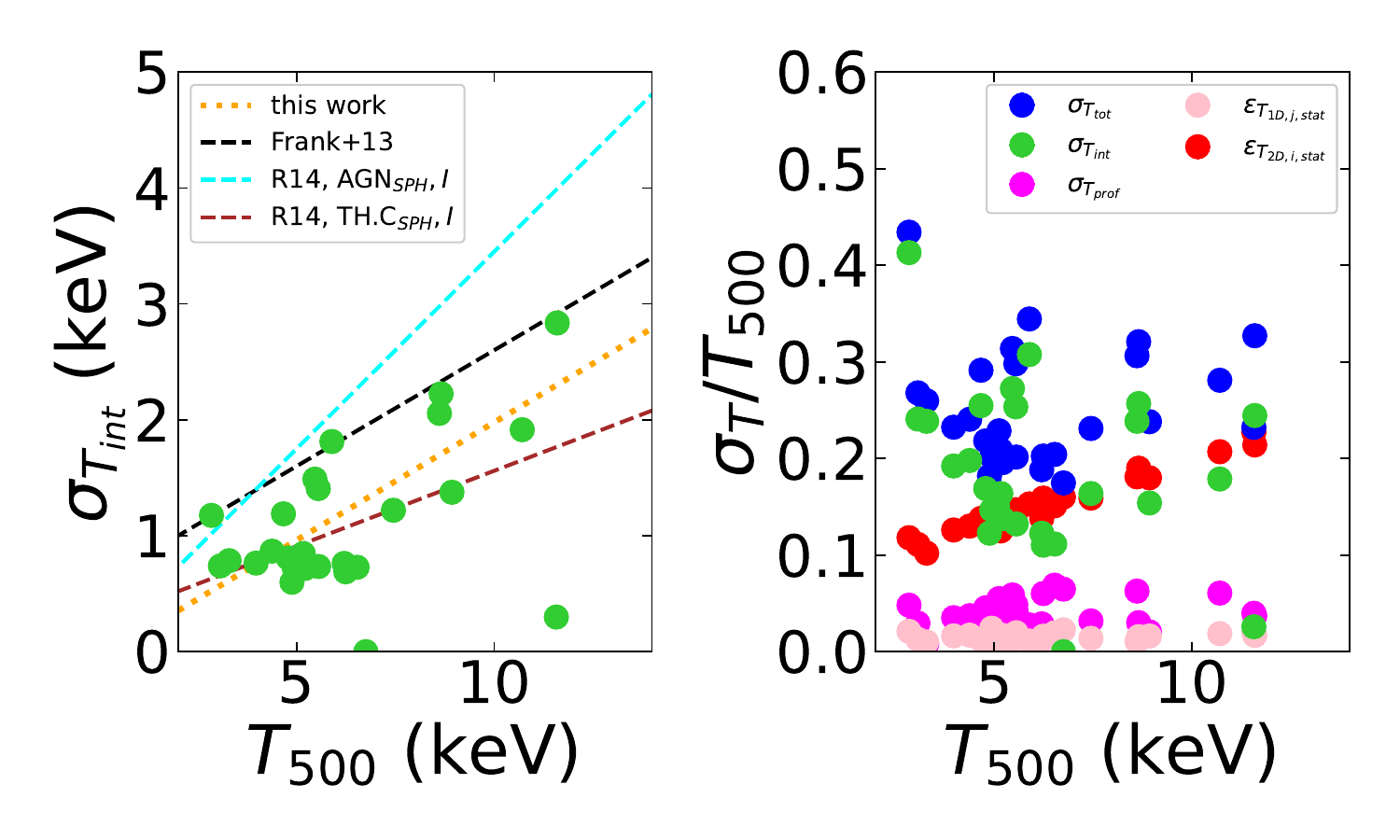}
\caption{
Total temperature dispersion within $R_{500}$ as a function of the overall cluster temperature.
{\it Left:} relation between the mean cluster temperature $T_{500}$ and the intrinsic scatter $\sigma_{T_{int}}$ within $R_{500}$. The dotted orange line represents the best-fit relation to the measured points and is compared to the observations by \cite{Frank2013} (black) and simulations with (green) and without (cyan) thermal conduction by \cite{Rasia2014}. {\it Right:} relative scatter as function of the cluster temperature. In blue, we show the observed values (i.e., total scatter $\sigma_{T_{tot}}/T_{500}$), in green we plot the relative intrinsic scatter (i.e., $\sigma_{T_{int}}/T_{500}$) computed as given in Eq. \ref{eq:sigmaTint} where the statistical errors (i.e, $\epsilon_{T_{1D,j,stat}}$) are shown in pink and are computed using Eq. \ref{eq:epsilon}, in magenta we show the scatter associated with the underlying temperature profile (i.e., $\sigma_{T_{prof}}/T_{500}$), and in red the average statistical uncertainty of the $T_{2D,i}$ measurements (i.e, $\epsilon_{T_{2D,i,stat}}$).
}
\label{fig:sigT}
\end{figure*}

\subsection{Global intrinsic scatter}
\label{sect:globalscatter}
In the previous section, we discussed the radial scatter (i.e., the dispersion in different annuli with size 0.1$R_{500}$). 
Here, we discuss the total dispersion within $R_{500}$ (i.e., the dispersion of all the cells around the temperature of the cluster, $T_{500}$). 
On top of the intrinsic variation of the ICM temperature distribution (due to turbulent motions and mergers) and the statistical uncertainties, we have also to account for the contribution due to the intrinsic radial variation over the same region. In fact, since the ICM is not isothermal, an underlying temperature profile will result in a distribution of spectroscopic measurements over the map with a given dispersion. 
Therefore, we compute the true intrinsic dispersion in the $kT$ distribution as 
\begin{equation}\label{eq:sigmaTint}
    \sigma_{T_{int}} = \sqrt{\sigma_{T_{tot}}^2 - \sigma_{T_{prof}}^2 - \epsilon_{T_{1D,j,stat}}^2 - \epsilon_{T_{2D,i,stat}}^2 } ,
\end{equation}
where $\sigma_{T_{tot}}$ and $\epsilon_{T_{1D,j,stat}}$ are computed using Eq.~\ref{eq:fluc_sigma} and Eq.~\ref{eq:epsilon} replacing $T_{2D,j}$ with the overall cluster temperature $T_{500}$, and  $\epsilon_{T_{2D,i,stat}}$ is the average statistical uncertainty of the $T_{2D,i}$ measurements. 
For each object with a given $M_{500}$ and redshift (see Table \ref{tab:clusters}), we recovered the expected projected temperature profile, $\sigma_{T,prof}$, by assuming the universal profile in \cite{ghi19_univ}.

In Fig.~\ref{fig:sigT}, we show that most of the scatter in the maps is associated with real inhomogeneities in the gas and not with the dispersion caused by the underlying profile or to the statistical uncertainties. Our results are in quite good agreement with the finding by \cite{Frank2013}, although with moderately lower intrinsic scatter values.
The fact that there is a significant scatter in the $\sigma_{T_{int}}$--$T$ relation may suggest that the specific details of the cluster history (e.g., mergers, gas cooling) play a significant role in the temperature distribution. 
The measured scatter is weakly dependent on the number of cells in the maps (r=0.36, p=0.06).
The positive correlation observed in Fig.~\ref{fig:sigT} ({\it left panel}) may suggest that the stronger shocks and turbulence of the more massive systems are driving the large range in values of the scatter. 
However, the most massive clusters are also those where we expect the largest impact of substructures because of the large differences between the temperature of the infalling component with respect to the temperature of the main cluster.
Nevertheless, the relative intrinsic scatter is much less dependent of the overall temperature (i.e., $r\sim$-0.16, p=0.44; see also {\it right panel} of Fig.~\ref{fig:sigT}) and, has values ranging between 0.11 and 0.41 (excluding the two clusters for which the statistical errors prevents a proper determination of the intrinsic scatter), suggesting that the relative intrinsic scatter is almost independent of the mass (confirmed by the low value of the Spearman coefficient, $r=-0.12$, p=0.58). 
The relative intrinsic scatter values are also independent of the dynamical state (i.e. correlation with $M_{all}$ gives $r=-0.00$, p=0.98).

\section{Discussion}
\label{sect:dis}

Gas inhomogeneities are ubiquitous in the ICM and affect various scales \citep[see, e.g.,][]{nag11,ron13,vaz13,Towler2023}. Hydrodynamical simulations show that, even for relaxed clusters, the electron density in a given radial shell can be well described by a log-normal distribution plus a possible tail (\citealt{zhu13}).
The former component, which accounts for a large fraction of the volume,  can be considered as nearly hydrostatic and can be used, for instance, to infer the total hydrostatic mass. 
A similar effect is present in the temperature distribution. It is thus crucial to understand and model the temperature inhomogeneities for a better characterization of the temperature profile. 
In fact, if multi-phase components coexist with the virial temperature, then a spectroscopic measurement, obtained assuming a  single-temperature thermal model, can be biased low with respect to the ``physically-motivated'' gas mass-weighted temperature within the same radial shell.

Our results show that a log-normal distribution is a good description of the 2D spectroscopic measurements of the temperatures in the ICM for most of the clusters. 
There are hints for a tail of high-temperature values as observed for the electron density. 
This is in agreement with numerical simulations (see, e.g., \citealt{Kawahara2007,zhu13}) which however often use 3D information. 
It is worth noticing that the reconstruction of the 3D distribution from observations is limited by the geometry of the gas halo and from projection effects (significantly unknown for disturbed clusters). 

Since a disturbance in the ICM is expected to contribute to the observed fluctuations and to the estimated scatter we explored if spectroscopic parameters (e.g., $\overline{|s_i|}$, std($s_i$), $\overline{\sigma_{T_{i,int}}/T_j}$) derived from the temperature maps can be used as a quantitative measure of the dynamical state of clusters, similar to what is done by morphological parameters as the surface brightness concentration and centroid-shift (among the most robust morphological parameters; see \citealt{lov17,cam22}). 
Our results provide no evidence for such correlation, possibly because the amplitude of the clumps in the ICM temperature is somewhat smaller than the ones in gas density, but also because the current resolution does not allow us to identify small features. 
Despite the lack of correlation between the spectrally derived parameters and the standard morphological estimators, in some cases, the spectral information can still provide complementary information about the dynamical state of clusters. 
A good example is G266.04-21.25 (aka ``Bullet'' cluster): the standard morphological parameters are unable  to identify it as a disturbed system (e.g., $M_{all}$=0.02); on the contrary, the values of the spectroscopically derived parameters are among the highest observed in our sample (see Table \ref{tab:specpar}).  

The measurements of the radial behavior of the temperature and density fluctuations in our sample 
suggest that there is a mix of isobaric and adiabatic fluctuations for most of the clusters (see Fig. \ref{fig:scatterR}). 
For comparison, the temperature and density gradients measured by \cite{sch04} in Coma suggest that the substructures are closer to the adiabatic case (see also \citealt{2016ApJ...818...14A}), while the finding by, for instance, \cite{Zhuravleva2018} suggests that the perturbations in the core are mainly isobaric.
By comparing perturbations within and outside 100 kpc \cite{hof16} found an indication for a change from isobaric to adiabatic perturbations.
Since our maps cover the full volume within $R_{500}$ (although  most of the cells are within 0.5$R_{500}$), our study of the temperature fluctuations is in agreement with previous studies and may confirm that isobaric processes dominate the core while adiabatic processes are often localized in the outskirts. 

The observed temperature inhomogeneities affect the global and radial temperatures which in turn can affect the mass bias level.  
\cite{ras12} showed that in simulations $\sim$10-15\% of the total mass bias can be attributed to such inhomogeneities, therefore by removing them we may indeed reduce the mass bias. 
Because of the current quality of the data, we could only show the impact of removing 1$\sigma$ level inhomogeneities (Fig. \ref{fig:Tclip}). 
With deeper exposures, providing smaller statistical uncertainties, we can relax such strong threshold.   
It is useful to note that since the observed inhomogeneities affect the derived integrated properties, these results not only may provide complementary information about the dynamical state of a system (e.g., the Bullet cluster), but, more importantly, can provide a statistical tool to evaluate how  fluctuations impact the general behavior of the relations among these integrated quantities. 
The estimated level of fluctuations, like the scatter of their distribution, should be included in the statistical analysis for a better understanding of the scaling laws. 

\begin{figure}[t]
\centering
\includegraphics[width=0.5\textwidth]{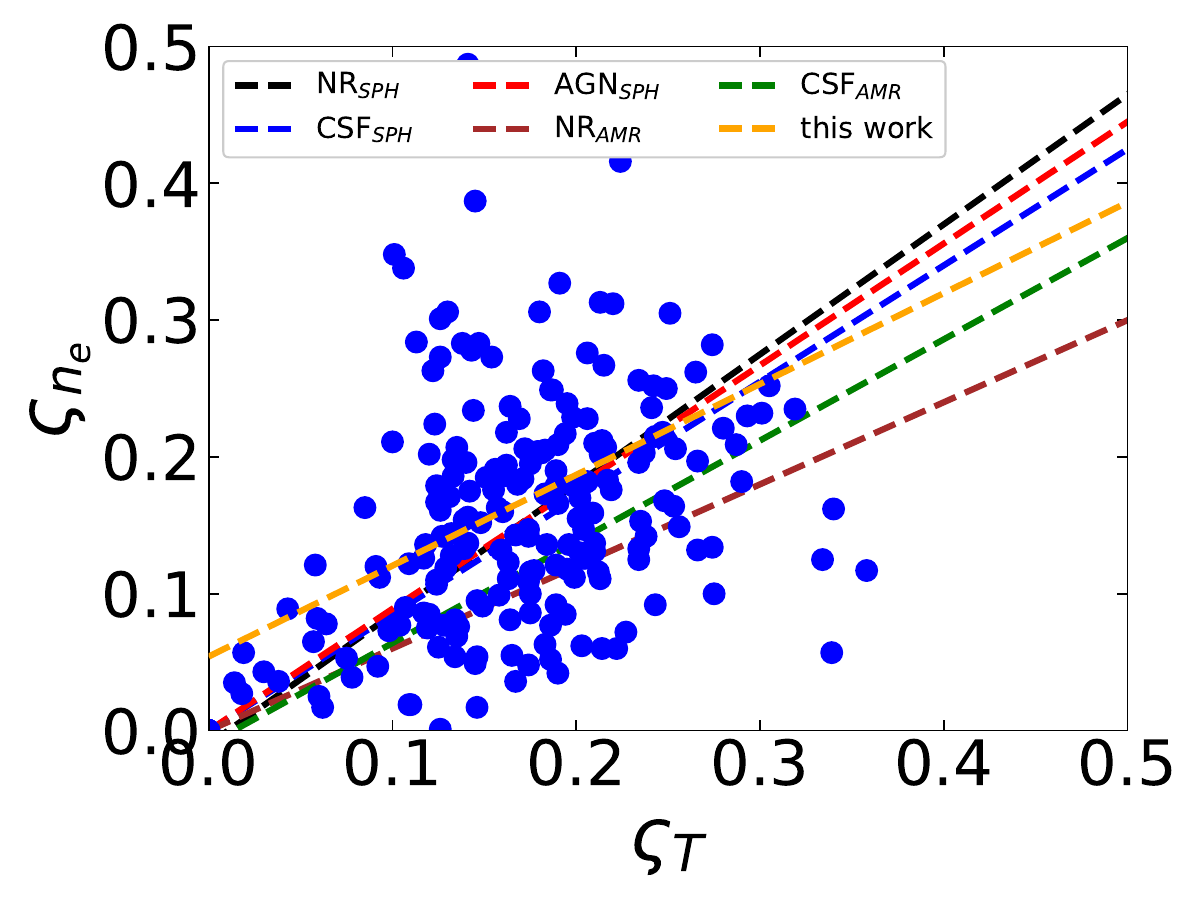}
\caption{
Correlation between Gaussian standard deviations $\varsigma_T$ and $\varsigma_{n_e}$ measured in different radial shells. Each data point corresponds to the values from a single shell of a cluster. The orange line represents the best-fit relation of the data points and is compared with the best-fits obtained by \cite{Rasia2014}. 
}\label{fig:sigmaTne}
\end{figure}

\subsection{Comparison with hydrodynamical simulations}
\label{sect:hydrosim}

By comparing the level of temperature inhomogeneities in our cluster sample with those reconstructed in objects extracted from hydrodynamical simulations, we can give insights into the physical mechanisms at work in the ICM. 
In particular, we compare our results with those obtained from different simulations (FLASH, see \citealt{gas14}; The Three Hundred carried out with the GADGET code, The300-GX -- see \citealt{cui2018}; and The Three Hundred carried out with the GIZMO code, The300-GIZMO -- see \citealt{cui2022}) and different physics. The cosmological simulations include the structure assembly and inflows, but have low resolution.
Hydrodynamical simulations test pure turbulence without substructures but have high resolution (i.e., a few kpc). 

Following \cite{Rasia2014}, we calculate the logarithmic gas density and temperature distributions in equal radial shells with a size of 0.1$R_{500}$. 
We refer to $\varsigma_{ne}$ and $\varsigma_{T}$ as the standard deviations of Gaussian distributions of the logarithmic values. Apart from some differences in smoothed-particle-hydrodynamics (SPH) and adaptive-mesh-refinement (AMR) codes (see discussion in \citealt{Rasia2014} for all the ICM physics prescription, including non-radiative / NR, cooling-star-formation without / CSF, and with BH growing and AGN feedback / AGN) the agreement between observation and simulations is quite good (see Fig. \ref{fig:sigmaTne}). 
For the values of $\varsigma_{ne}$ and $\varsigma_{T}$ covered by our observations, it seems that the agreement is better with SPH simulations which are characterized by a higher level of temperature fluctuations. 

\begin{figure}
\centering
\includegraphics[width=0.5\textwidth]{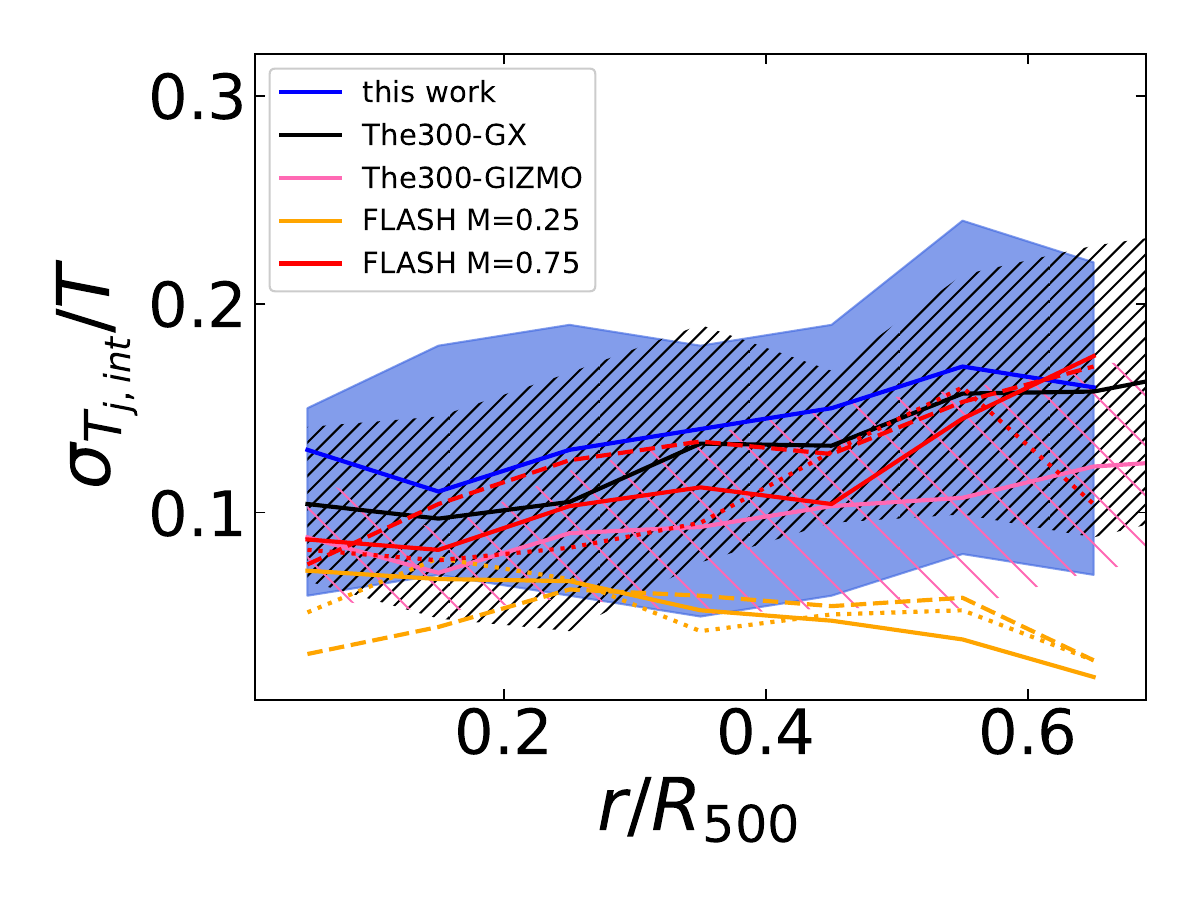}
\caption{
Comparison of $\sigma_{T_{j,int}}/T$ estimated in cells of $0.1 R_{500}$ as a function of radius between the CHEX-MATE objects (blue; the 16th and 84th percentiles were computed excluding the annuli for which the statistical errors prevents a proper determination of the intrinsic scatter) and results extracted from various hydrodynamical simulations (see text for details in Sect.~\ref{sect:hydrosim}). For cosmological simulations (i.e., The300 GX and GIZMO), we represent the distribution for 45 objects with the median and a dispersion equal to the inter-quartile range divided by 1.35. 
For the FLASH simulations, we present the results obtained for 3 different projections (with different line styles) of the same object  simulated with two different levels of turbulence (i.e., different Mach).}
\label{fig:comp2sim}
\end{figure}

In Fig.~\ref{fig:comp2sim}, we compare the radial behavior of the ratio between the intrinsic dispersion and the median value of the ICM temperature as obtained for this CHEX-MATE subsample and a set of different hydrodynamical simulations.
We also verified the impact of the Voronoi tesselation (specifically, we produce temperature and emissivity maps from the model and use the Voronoi binning adopted in the spectral analysis to recover the expected contribution of the model on $\sigma_{T_{j,int}}/T$) and found that is negligible.
There is a general agreement on the trend with the radius between the constraints obtained by observations and simulations, with a very weak increase ratio moving outwards. There is also a good agreement in the level of the temperature dispersion measured in observations and simulations. 
We also note that, in constrained hydrodynamical simulations, a relatively high turbulence level (but still subsonic) is required to reach values closer to the observed data (and to the results of the cosmological simulations).

High-resolution simulations have shown a connection between such fluctuations and the Mach number of gas motion in the ICM (e.g., \citealt{gas13,gas14}). 
In the low Mach number regime (i.e., $\Mach$<0.5) perturbations are mainly isobaric and therefore one expects $\sigma_{T}/T \sim \sigma_{n_e}/n_e$, while for high Mach numbers (i.e., 0.5<$\Mach$<1) the perturbations shift to the adiabatic regime implying (e.g. for $\gamma$=5/3) $\sigma_{T}/T \sim 0.67 \sigma_{n_e}/n_e$. 
The fluctuations in the ICM move from isobaric to adiabatic from the $\Mach_{3D}$=0.25 to $\Mach_{3D}$=0.75.  
This is true for subsonic turbulence but a compressional components may also lead to weak shocks and sounds waves (i.e., adiabatic fluctuations) even in low-Mach case (e.g., \citealt{Mohapatra2022}). 
However, in a seminal work, \cite{Ryu2008} showed that turbulence in clusters is largely solenoidal with subsonic velocities (see also \citealt{Miniati2014,vaz2017}).
We also note that the low/high turbulence simulations by \cite{gas13} and \cite{gas14} mimic a relaxed/unrelaxed cluster (or merger-like vs. internal turbulence). 
The temperature features become more washed out in the simulated clusters with lower turbulence (i.e., relaxed system). 
Even though in the relaxed clusters shocks are weak (essentially sound waves, as shown by the absence of thin sheets), some extended rolls/eddies/rarefactions are still visible in the spectroscopic-like temperature map.

\subsection{Role of the thermal conduction in the ICM}

The level of ICM inhomogeneity may also depend on diffusive processes, such as thermal conduction. 
In fact,  thermal conduction  makes the gas more isothermal by smoothing out the temperature substructure in the ICM and therefore reducing the values of $\sigma_T$. 
This should be particularly true at high temperatures where conductivity, being strongly temperature dependent ($k\propto T^{5/2}$), becomes more efficient (e.g., \citealt{Dolag2004}). 
Thus, the slope of the $\sigma_T$--$T$ relation can be used to qualitatively constrain the degree of thermal conduction in the intracluster plasma. 
The observed relation stands between the values obtained from cosmological simulations with and without thermal conduction using a conductivity of 1/3 of the Spitzer value (see Fig. \ref{fig:sigT}). 

The link between thermal conduction and turbulence is particularly important, as turbulence can re-orient the magnetic field and thus reduce or restore heat flow to a region.
Thermal conduction has been mainly investigated via theoretical studies (e.g., \citealt{Dolag2004,ZuHone2013,gas14,Biffi2015}) because, from an observational point of view, it is very challenging to resolve local features. 
However, the observations of sharp temperature gradients linked to cold fronts (\citealt{Ghizzardi2010,ZuHone2013}) and the presence of cool cores favor a scenario where the conduction is highly suppressed (\citealt{Molendi2023}, and refs therein). 
Our results suggest that the level of thermal conductivity, although non-zero, is smaller than 1/3 of the Spitzer value (i.e., the value assumed in the cosmological simulations used for the comparison).  
The low value of thermal conduction is also suggested by the hydrodynamical simulations. 
In fact, the red lines in Fig. \ref{fig:comp2sim} which are obtained with a relatively high Mach number  are already unable to fit the observed data. Any extra conduction will further suppress the scatter down to lower values requiring even higher Mach numbers to match the observed trend.

\subsection{Connections with turbulence and mass bias $b$}
\label{sect:eturb}

Our work focuses on the variations we resolve in maps of the temperature measurements that we obtain from the fitting of counts-based spectra, integrated along the line of sight.
We discuss here the physical implications of such fluctuations in the spectral temperature distribution. 
By interpreting most of these fluctuations as generated from turbulence, induced by the mass accretion in the process of cluster formation, we can infer the ratio between the energy in turbulence and the thermal energy, and translate this ratio in terms of a predicted value of the hydrostatic mass bias $b = 1 - M_{\rm HE}/M_{\rm tot}$ \citep[see, e.g.,][]{kha16,ett22}.

\begin{figure}
\centering
\vbox{
\includegraphics[width=0.5\textwidth]{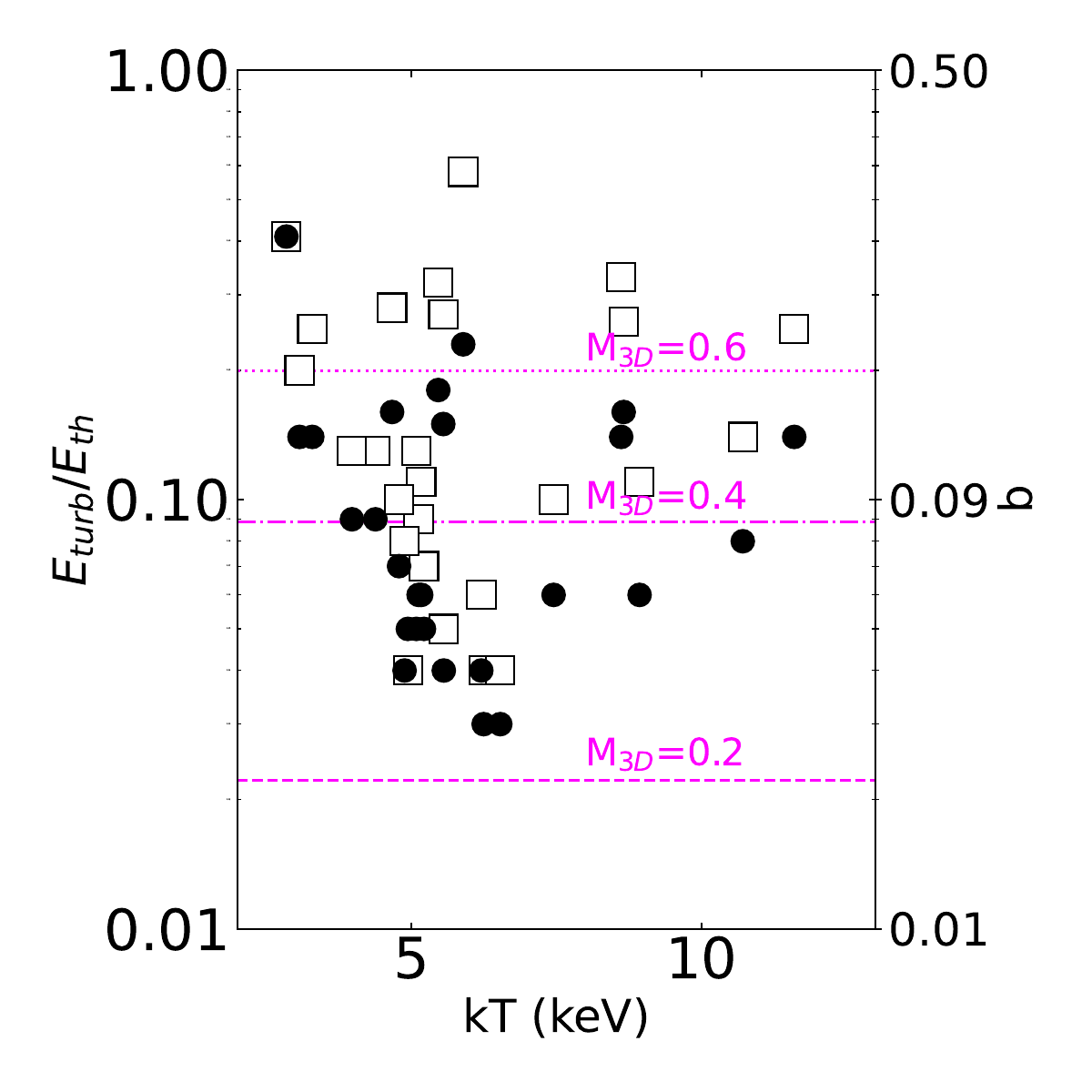}
\includegraphics[width=0.40\textwidth]{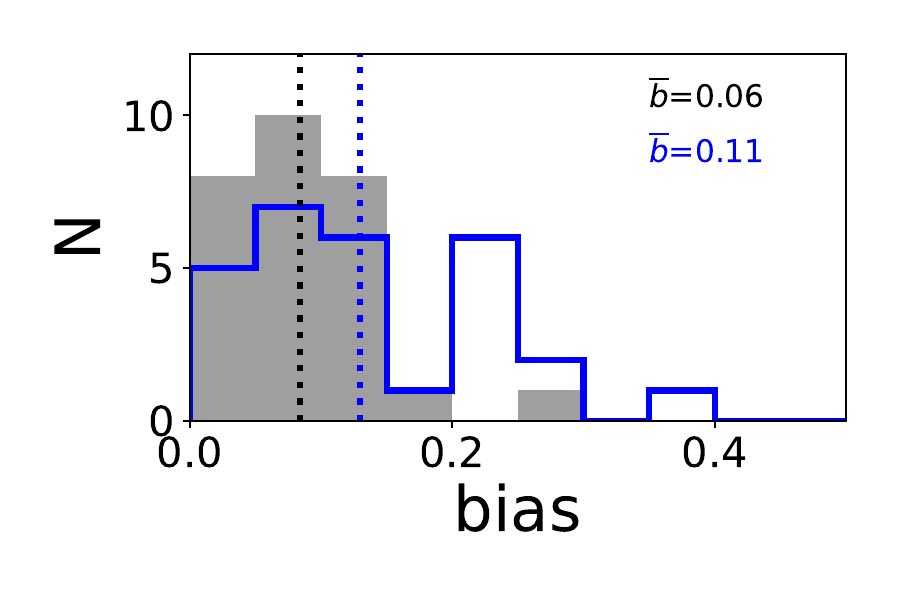}
}
\caption{
Mass bias and ratio between turbulent and thermal energy in the clusters in our sample.
{\it Top:} ratio between the turbulent and the thermal energy computed based on the $\sigma_{T,int}/T$ values and under the assumption that the isobaric fluctuations are dominant.  The dashed horizontal lines are computed for different Mach numbers using the following Eq.: $E_{\rm turb}/E_{\rm th} = 0.5 \, \gamma \, (\gamma-1) \, \Mach_{3D}^2$.  In the right y-axis, we provide the mass bias $b$ estimated within $R_{500}$ through Eq.~ \ref{eq:bias}. The empty squares indicate the values of $E_{turb}/E_{th}$ (and $b$) after the correction accounting for the underlying power spectrum (see the text for more details).
{\it Bottom:} distribution of the values of the hydrostatic bias. In gray (blue), we show the $b$ values before (after) accounting for the integration of the power spectrum between some characteristic scales (see text in Sect.~\ref{sect:eturb}). 
Dotted lines indicate the means of the distributions.
}
\label{fig:EturbEth}
\end{figure}

In the adiabatic regime, we have $\sigma_T/T \sim 0.67 \sigma_{n_e}/n_e$, while in the isobaric regime $\sigma_T/T \sim \sigma_{n_e}/n_e$ (see \citealt{gas14}). 
Since our results suggest a mix of the two regimes (see Fig. \ref{fig:scatterR}), in first approximation we can take the mean of them, i.e, $\sigma_T/T \sim 0.83\sigma_{n_e}/n_e$. 
We can now estimate the Mach number by applying the relation  $\Mach_{1D}\sim\sigma_{n_e}/n_e\sim1.2*\sigma_T/T$ which gives  $\Mach_{3D}= \sqrt(3)*\Mach_{1D}\sim2.1*\sigma_T/T$.
We measured $\sigma_{\rm T,int}/T$=0.17$^{+0.08}_{-0.05}$ (see Sect. \ref{sect:globalscatter} and Fig. \ref{fig:sigT}). 
Therefore, $\Mach_{3D}=0.36^{+0.16}_{-0.09}$. Hydrodynamical simulations of galaxy clusters usually find $E_{\rm turb} / E_{\rm th}$ in the range 5-50\% (e.g., \citealt{vaz2009,lau2009,gas2012}). 
For $E_{\rm turb}/E_{\rm th} = 0.5 \, \gamma \, (\gamma-1) \, \Mach_{3D}^2$  that translates into a $\Mach_{3D}$ from simulations in the range 0.30-0.95, consistent with our observational constraints. 
Our result is also in agreement with the finding of \cite{hof16} who found $\Mach_{3D}$=0.16-0.40.  
It is worth noticing that since we do not remove the substructures it is possible that the measured temperature fluctuations are not only associated with turbulence. 
Therefore, the derived constraints should be considered as upper limits.
However, given that the measured temperature scatter does not correlate with the dynamical state of the systems, the presence of significant substructures (which is true only for a very few systems in our sample) may not be playing a significant role.  
Moreover, if cosmological substructures were to dominate, $\Mach$ numbers would appear supersonic (mimicking shocks) in the majority of systems because of the substantial boost in the density maps.

The measured $\Mach_{3D}$ corresponds to $E_{turb}/E_{th}\sim0.07^{+0.09}_{-0.03}$ which is in agreement with the observational finding of \cite{eck19} and \cite{2023A&A...673A..91D}, and of the theoretical work by \cite{Angelinelli2020}. 
As discussed in \cite{ett22} \citep[see also][]{kha16}, we can then relate the quantity $E_{\rm turb} / E_{\rm th}$ to the hydrostatic mass bias $b$ as 
\begin{equation}\label{eq:bias}
    b = \left( \frac{E_{\rm th}}{E_{\rm turb}} +1 \right)^{-1}.
\end{equation}

In Fig.~\ref{fig:EturbEth}, we show the estimated ratios between the turbulent and thermal energy within $R_{500}$ which points, for the studies sample, to a mass bias of $b$=0.06$^{+0.07}_{-0.03}$, comparable to that derived by the X-COP collaboration for a small sample of massive, nearby, mostly relaxed clusters \citep[see, e.g.,][]{ett19,eck19}. 

The variance in the maps is related to the turbulent energy $E_{turb}$, which is proportional to the integral of the power spectrum $P(k)=k^{\alpha}$:
\begin{equation}\label{sigmaPk}
\sigma_T^2\propto E_{turb}\propto\int{P(k)k^2dk}=\int{E(k)dk},
\end{equation}
with the energy spectrum $E(k)=k^{\alpha+2}$. Eq. \ref{sigmaPk} is fully generic and can be applied to any mode of fluctuation. 
For Kolmogorov-like turbulence (i.e., $\alpha$=-11/3 leading to $E(k)\propto k^{-5/3}$), we expect $\sigma_T$ to increase at larger scales (i.e., for decreasing $k$).
In our calculations, being the dispersion measured on some characteristic scales regulated by the resolution of the maps and the overall volume sampled, we have to estimate a correction to $E_{\rm turb}$ by integrating the power spectrum between a dissipation scale and an injection scale. We take these to be 10 kpc and $0.5R_{500}$ (see \citealt{gas14}), respectively. 
The shape of the power-spectrum is assumed to be Kolmogorov-like. With the correction for the power spectrum we obtain $E_{turb}/E_{th}\sim0.12^{+0.16}_{-0.08}$ and $b\sim$0.11$^{+0.11}_{-0.07}$ (see Fig.~\ref{fig:EturbEth}) 
which tend to agree better to what derived from X-ray vs lensing studies (e.g., \citealt{Sereno2015,her2020,lov2020}), and to recent hydrodynamical simulations (e.g., \citealt{2021MNRAS.506.2533B,2023MNRAS.518.4238G}).
It is worth noticing that the injection and dissipation scales from simulations are not well known and depends on sub-grid models.

\section{Conclusions}
\label{sect:con}

We investigate the level of inhomogeneities in the gas temperature (and gas density) 2D distribution for a sample of 28 CHEX-MATE galaxy clusters. 
The temperature maps show clear structures that we associate with fluctuations in the ICM of heterogeneous origin (mostly due to the ongoing accretion processes of, for instance, cold clumps and subhalos, and energy diffusion related to the feedback). The distributions of the temperature bins are well described by a log-normal function (see Fig.~\ref{fig:skewQQ}).
Once these inhomogeneities are removed, the reconstructed temperature profiles vary, with the intensity of the variation increasing with the radius (see Fig. \ref{fig:Tclip}). That has indeed the effect of changing the absolute temperature and gradient of the profiles with an obvious impact on the estimate of the fundamental integrated properties (e.g., the total mass).

The overall level of turbulence, resolved in our spectral analysis, suggests
that, on average, there is a mix of isobaric and adiabatic fluctuations (see Fig.~\ref{fig:scatterR}). We constrained the average 3D Mach number in our sample to $\Mach_{3D}$=0.36$^{+0.16}_{-0.09}$. The global effect translates into a level of energy in turbulence that is about 7\% of the thermal energy (see Fig.~\ref{fig:EturbEth}), implying an estimated hydrostatic bias of approximately 6\% (ranging between 0 and 29\%) in this representative subsample of CHEX-MATE clusters.
However, when we apply the correction to $E_{turb}$ by integrating the power spectrum between a dissipation scale and an injection scale, we obtain a median hydrostatic bias of $b\sim$11\%, covering a range between 0 and 37\%.

Once resolved as a function of the radius, the estimates of $\sigma_{T_{j,int}}$/$T$ match the ones from state-of-art hydrosimulations (see Fig. \ref{fig:comp2sim}). 
Also, the observed $\sigma_{T_{int}}$-$T$ relation is slightly steeper than the one obtained with hydrodynamical simulations including thermal conduction (which drastically smooths temperature variations and homogenizes the ICM; see Fig.~\ref{fig:sigT}) at 1/3 of the Spitzer value, but it is significantly shallower than the one derived without thermal conduction. 

The knowledge acquired by this work on the level of fluctuations present in the gas temperature maps, together with the evaluation of the dynamical state obtained by the estimate of the morphological parameters determined from the analysis of the X-ray images \citep{cam22}, will allow us both to guide the interpretation on, and to quantify, the amount of scatter that will be resolved both in the thermodynamic radial profiles and in the integrated quantities of the CHEX-MATE clusters. 
Given the encouraging results presented in this work, the next step will be to extend this kind of analysis to the entire CHEX-MATE sample and to investigate the link between the
observed gas $T$, gas density, and surface brightness fluctuations to the scatter characterizing the distribution of the integrated quantities in the scaling relations (e.g., \citealt{Pratt2009,LL2020}).
An investigation of the statistics of the fluctuations in the surface brightness maps of the CHEX-MATE objects, and relating the density fluctuations and the turbulent motions, will be presented in a forthcoming paper.

\begin{acknowledgements}
We thank the anonymous referee for her/his review of the manuscript.
We also thank M. Roncarelli for useful discussion.
L.L. and S.E. acknowledge financial contribution from the contracts ASI-INAF Athena 2019-27-HH.0, ``Attivit\`a di Studio per la comunit\`a scientifica di Astrofisica delle Alte Energie e Fisica Astroparticellare'' (Accordo Attuativo ASI-INAF n. 2017-14-H.0), and from the European Union’s Horizon 2020 Programme under the AHEAD2020 project (grant agreement n. 871158). 
L.L. also acknowledges financial contribution from the INAF grant 1.05.12.04.01. 
W.F. and C.J. acknowledge support from the Smithsonian Institution, the Chandra High Resolution Camera Project through NASA contract NAS8-03060, and NASA Grants 80NSSC19K0116, GO1-22132X, and GO9-20109X. 
B.J.M. acknowledges support from STFC grant ST/V000454/1.
G.W.P. acknowledges support from CNES, the French space agency.
J.S. and J.K. were supported by NASA Astrophysics Data Analysis Program (ADAP) Grant 80NSSC21K1571.
This research was supported by the International Space Science Institute (ISSI) in Bern, through ISSI International Team project \#565 ({\it Multi-Wavelength Studies of the Culmination of Structure Formation in the Universe}).
In this work we made use of the WVT binning algorithm by \cite{vor06}, which is a generalization of \cite{2003MNRAS.342..345C} Voronoi binning algorithm. We also made use of the following packages: FTOOLS (\citealt{1995ASPC...77..367B}), DS9 (\citealt{2003ASPC..295..489J}), Astropy (\citealt{2013A&A...558A..33A,2018AJ....156..123A}).

\end{acknowledgements}

\bibliographystyle{aa} 
\bibliography{2Dmaps}

\begin{appendix} 

\section{Gallery}
\label{gallery}

\begin{figure*}[h]\textit{\label{fig:gallery}}
\centering
\includegraphics[width=1\textwidth]{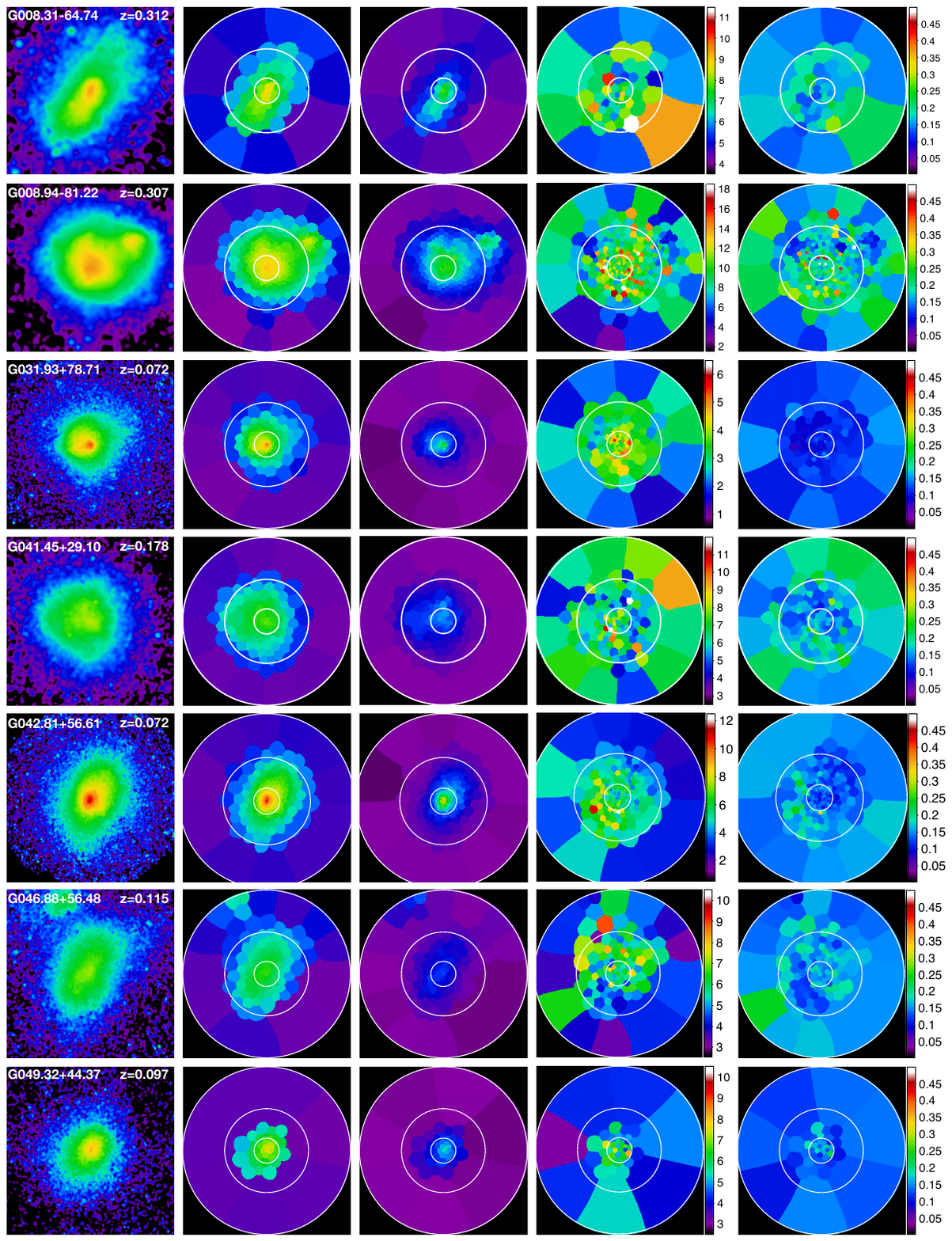}
\caption{
From {\it left} to {\it right:} (a) bkg-subctracted and exposure-corrected images in the 0.3-7 keV band; The size of the boxes corresponds to $R_{500}$; (b) binned surface brightness maps; (c) projected electron density maps (d) projected temperature maps; (e) relative temperature error maps. The white circles in the voronoi maps corresponds to 0.15$R_{500}$, 0.5$R_{500}$, and $R_{500}$.   }
\end{figure*}

In Fig. \ref{fig:gallery} we show the recovered surface brightness, electron density, and temperature maps within $R_{500}$ for all the clusters analyzed in this work. 

\begin{figure*}[t]
\centering
\includegraphics[width=1\textwidth]{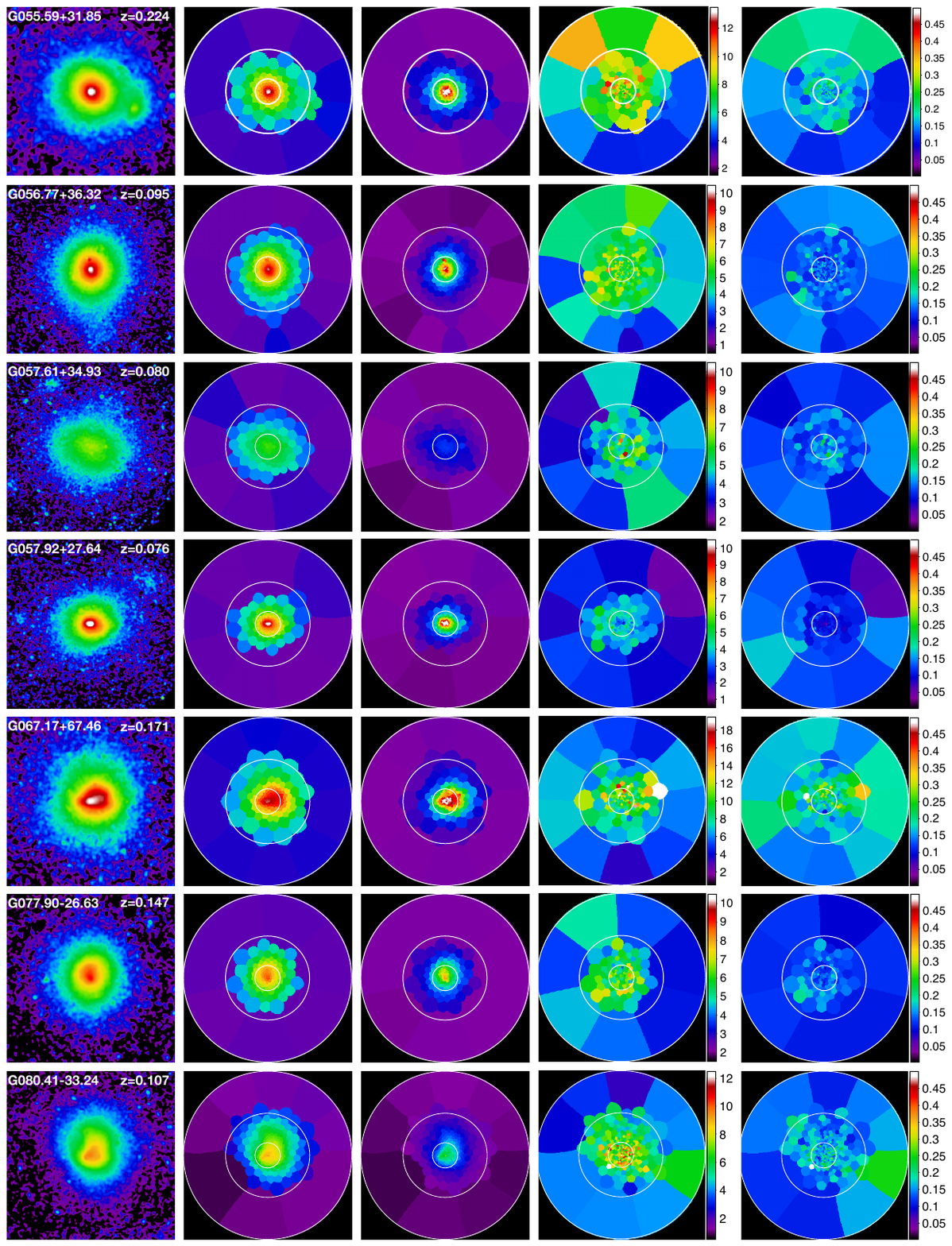}
\caption{
From {\it left} to {\it right:} (a) bkg-subctracted and exposure-corrected images in the 0.3-7 keV band; The size of the boxes corresponds to $R_{500}$; (b) binned surface brightness maps; (c) projected electron density maps (d) projected temperature maps; (e) relative temperature error maps. The white circles in the voronoi maps corresponds to 0.15$R_{500}$, 0.5$R_{500}$, and $R_{500}$.    }
\end{figure*}

\begin{figure*}[t]
\centering
\includegraphics[width=1\textwidth]{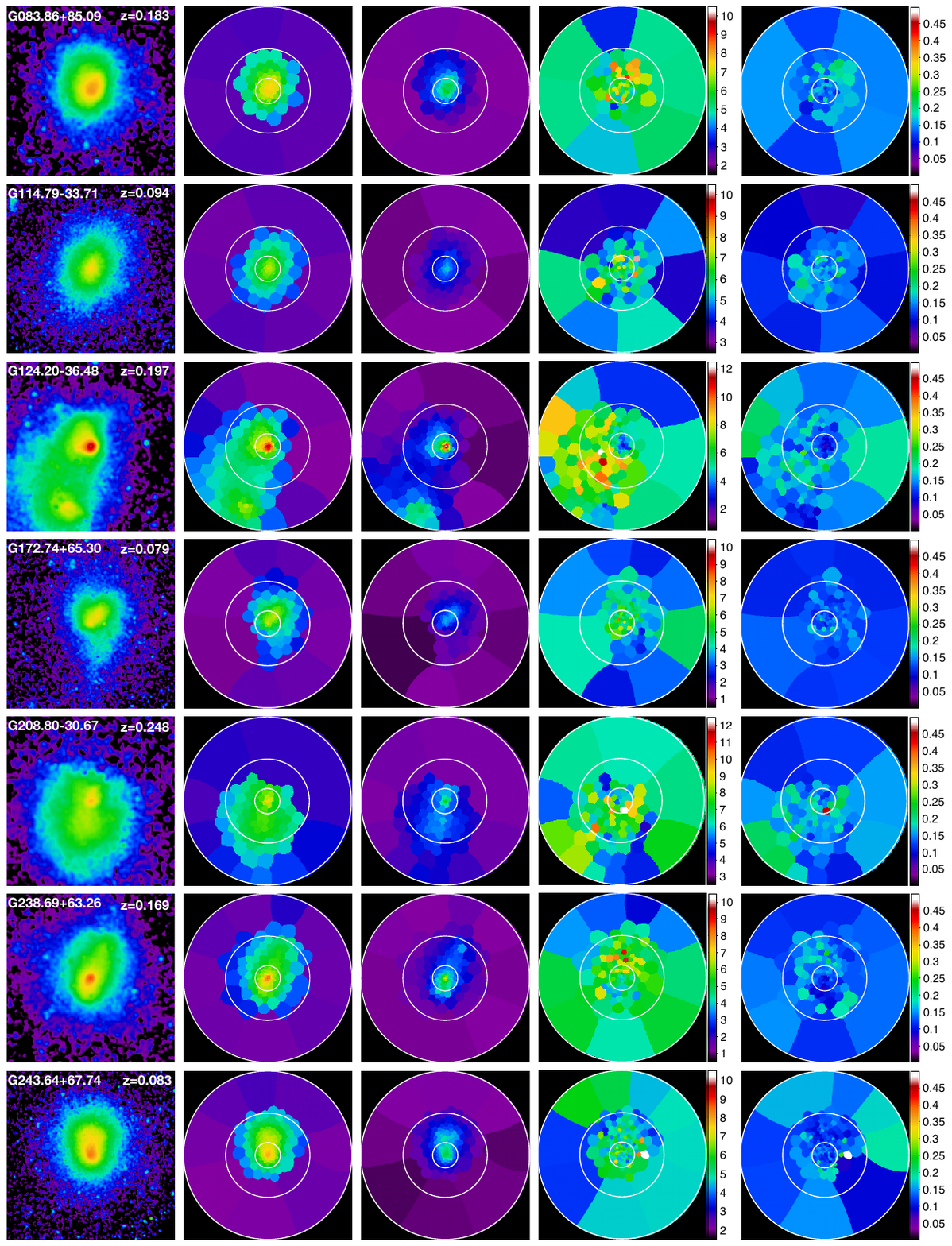}
\caption{
From {\it left} to {\it right:} (a) bkg-subctracted and exposure-corrected images in the 0.3-7 keV band; The size of the boxes corresponds to $R_{500}$; (b) binned surface brightness maps; (c) projected electron density maps (d) projected temperature maps; (e) relative temperature error maps. The white circles in the voronoi maps corresponds to 0.15$R_{500}$, 0.5$R_{500}$, and $R_{500}$.   }
\end{figure*}

\begin{figure*}[t]
\centering
\includegraphics[width=1\textwidth]{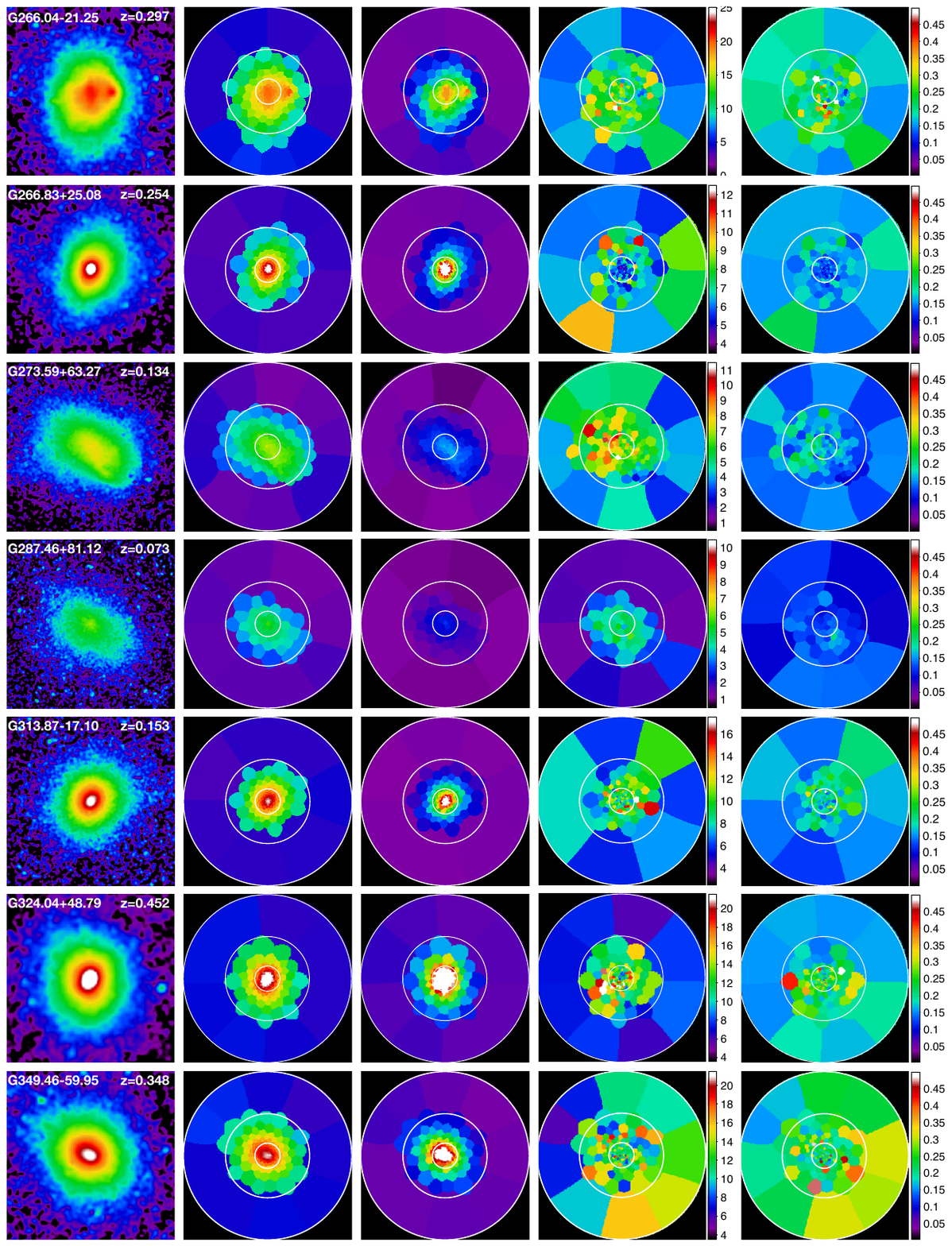}
\caption{
From {\it left} to {\it right:} (a) bkg-subctracted and exposure-corrected images in the 0.3-7 keV band; The size of the boxes corresponds to $R_{500}$; (b) binned surface brightness maps; (c) projected electron density maps (d) projected temperature maps; (e) relative temperature error maps. The white circles in the voronoi maps corresponds to 0.15$R_{500}$, 0.5$R_{500}$, and $R_{500}$.    }
\end{figure*}

\section{Comparison with the curvelet maps}\label{App:curvelet}

Thermodynamical 2D maps have been extensively used in the study of galaxy clusters, thanks to their great potential to characterize the dynamical state of a system. 
For this reason, a number of spectral-mapping methods have been developed. 
Based on the approach, these methods can be divided into three main categories: ({\it i}) hardness ratio method  which combines X-ray imaging in several energy bands (e.g., \citealt{Ferrari2006}); ({\it ii}) wavelet (see \citealt{Bourdin2004,Bourdin2008}) or curvelet (see \citealt{Bourdin2015}) analysis using the wavelet or curvelet coefficients to couple a multi-scale spectroscopic analysis with a structure detection scheme; ({\it iii}) spatially resolved spectroscopy of independent cells sampled by a required S/N or a minimum number of counts for which at least three different binning strategies have been proposed in this context: adaptive binning (e.g., \citealt{2001MNRAS.325..178S}, \citealt{LL09,LL2011}, \citealt{2011MNRAS.411.1833O}); contour binning (\citealt{Sanders2006}, \citealt{hof16}); Weighted Voronoi Tesselation (WVT) method (e.g., \citealt{2003MNRAS.342..345C}). In general, the Voronoi algorithms define the smallest possible regions for the spectral analysis, given the statistics of the observation. 
It is based on the distribution of the signal and noise in a given band and it puts together pixels to reach a required S/N, regardless of the ``temperature'' that would be measured in that cell. 
The wavelet or curvelet analysis has the advantage of being not driven by the distribution of the brightness of the emission. 
Conversely, the advantage of Voronoi algorithms is that (i) regions can be treated as independent from each other (when neglecting the effect of the PSF) and (ii) it does not try to follow the gradients of the ICM emission.  

\begin{figure}[h]
\centering
\includegraphics[trim=10 0 60 260, clip, width=0.45\textwidth]{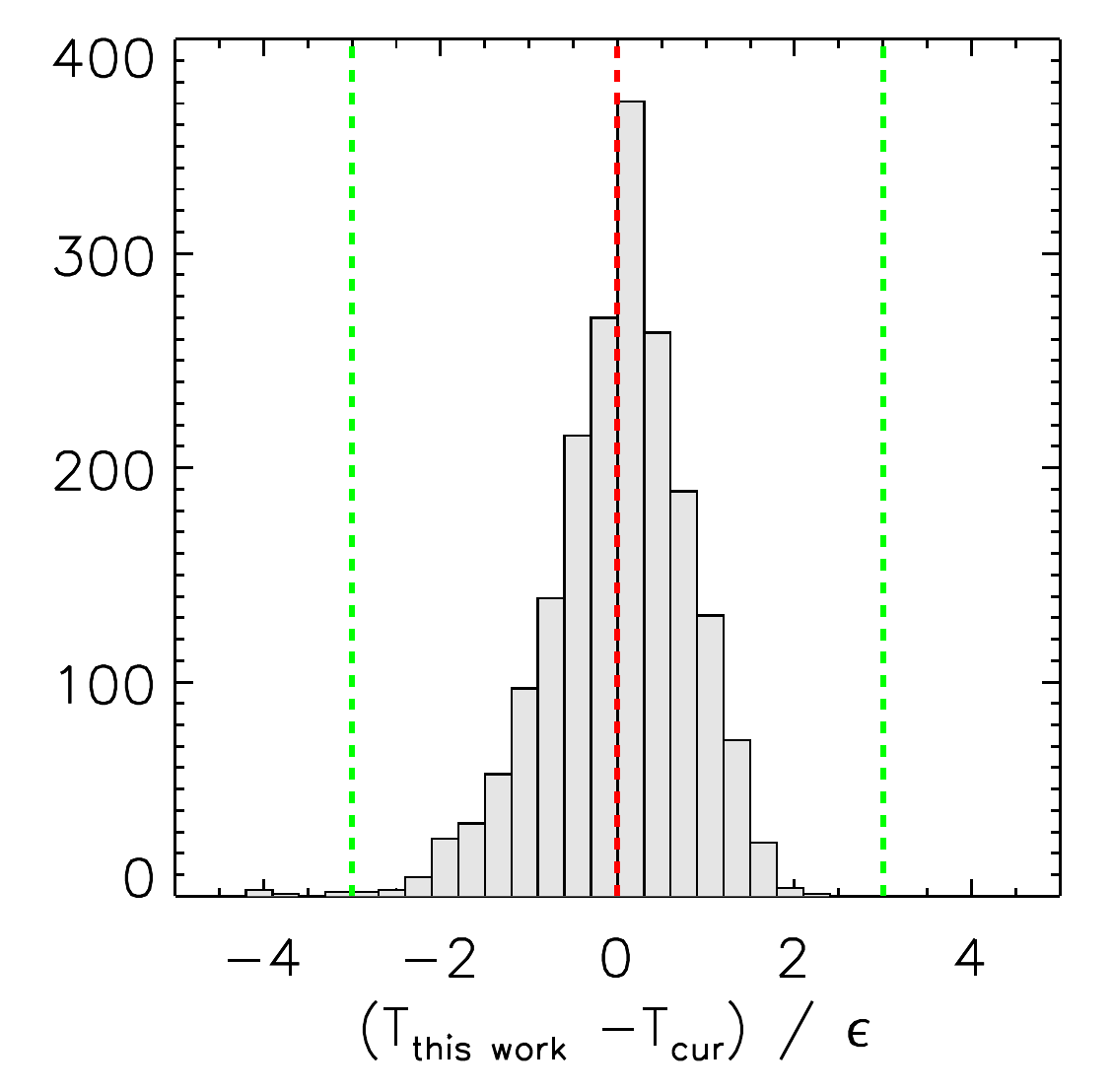}
\caption{Comparison of the temperatures measured in the maps reconstructed in this work and the ones obtained from the curvelet technique. The histogram represents the distribution of the difference in spectroscopic measurements normalized to the statistical error $\epsilon = \sqrt{\epsilon_{\rm this \, work}^2 +\epsilon_{\rm cur}^2}$ for 1,916 spatial cells obtained in 10 objects. The vertical dashed lines indicate 0 (red) and -3, 3 (green).}
\label{fig:curvelet}
\end{figure}

For ten objects in our sample, we compared the temperature map obtained through the spectral analyses of the Voronoi tessellation regions with the one obtained through the curvelet analysis. 
For each Voronoi region, we computed the weighted temperature from the curvelet map and we then compared the two $T$ values. 
In Fig.~\ref{fig:curvelet}, we show a summary plot to demonstrate that the temperatures recovered between the two methods are in good agreement, with an average value of the ratio $(T_{\rm this\, work} - T_{\rm cur})/\epsilon$ of 0.005 (median: 0.06; 1st-3rd quartile: -0.42, 0.53). 

We also compared the radial temperature fluctuations estimated from the Voronoi and curvelet temperature maps. 
We found a remarkably good agreement between the $\sigma_T/T$ recovered with the two methods in the regions between 0.2-0.8$R_{500}$. 
At large radii (i.e., $>0.8R_{500}$) the comparison is not straightforward because of the coarse mapping with the Voronoi method and of the masking of low signal-to-noise regions in the curvelet analysis. 
Within 0.2$R_{500}$, the $\sigma_T/T$ values in the Voronoi maps are slightly larger than what was found by the curvelet analysis. 
A detailed investigation is required to understand the cause of this slight difference but it is beyond the scope of the paper. 
However, this does not affect any of the main conclusions.

\section{Temperature cell weights}
\label{weights}
\begin{figure}[h!]
\centering
\includegraphics[width=0.5\textwidth]{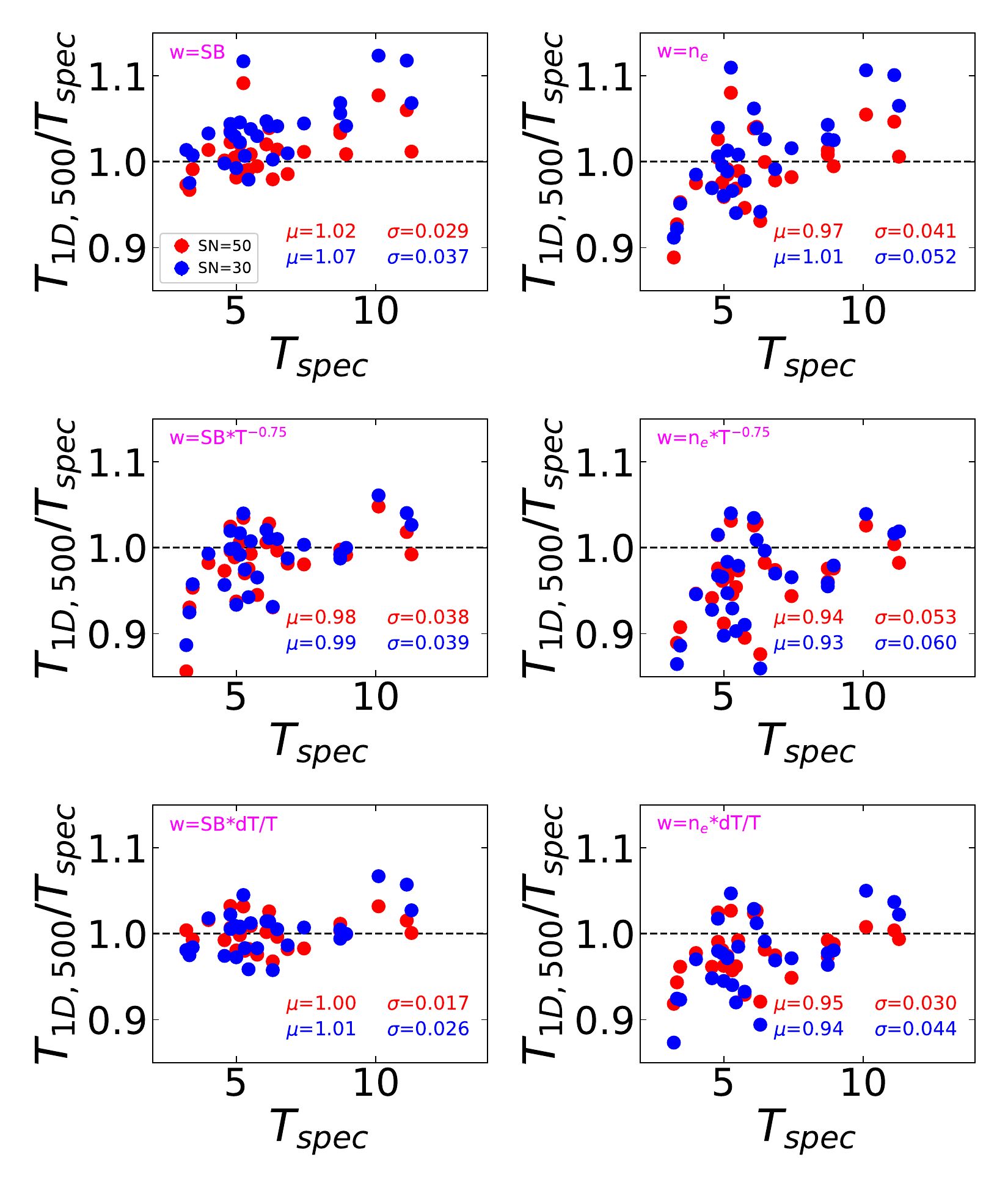}
\caption{
Comparison between the temperature derived from the maps (i.e., $T_{1D,500}$) and the global temperature (i.e., $T_{spec}$) obtained by fitting a single spectrum extracted within $R_{500}$. In each subplot the text in magenta indicates the weighting used to recover $T_{1D,500}$ from the maps, while the values of $\mu$ and $\sigma$ indicate the median of the distribution and its dispersion. In blue we show the results using the maps obtained with S/N=30 while in red we show the ones obtained with S/N=50. }
\label{fig:weights}
\end{figure}

In Fig. \ref{fig:weights}, we show how well the overall cluster temperatures recovered from the maps (i.e., $T_{1D,500}$) match the fitted values using one single extraction region (i.e.,  $T_{spec}$). 
We used six different weightings, each of them including a proxy of the X-ray emissivity (i.e., either the value of the surface brightness of the cells or the electron density obtained using Eq. \ref{eq:n_k}) multiplied by: (i) the area of the cell, (ii) the area of the cell multiplied by $T^{\alpha}$ (with $\alpha$=-0.75 from \citealt{maz04}), and (iii) the area multiplied by the relative errors of the cells. 
Since the relative errors are smaller for lower temperatures, both the latter approaches weight more the cells with lower temperatures.

In general, using SB instead of $n_e$ as a proxy for the X-ray emissivity returns $T_{1D,500}$ in better agreement with $T_{spec}$. This is possible because there is no assumption on the scale of the line of sight (i.e., the volume to be used to recover the electron density). If so, in principle this result could be used to provide a constraint on the elongation along the line of sight (helpful for a measure of the triaxiality), but this is beyond the scope of this paper. 

Using only the area of the cells as weights is clearly not sufficient (see {\it top panels} of Fig. \ref{fig:weights}) because it does not account for the energy dependence of the effective area. 
Using the relative errors of each cell in the weighting ({\it bottom panels}) returns  $T_{1D,500}$ in much better agreement with $T_{spec}$ (e.g., compare the $\mu$ and $\sigma$ values of the different subplots). Weighting each cell by $T^{\alpha}$ works also relatively well ({\it middle panels}), although there is an increasing deviation at lower temperatures.
However, since for comparison with the simulations using the factor $T^{\alpha}$ is the best option (e.g., \citealt{maz04}), and after verifying that the impact on the results of the paper is not significant, we used this latter weighting as our default method.

\section{Voronoi tessellation and map resolution}\label{multimaps}

\begin{figure}[!t]
\centering
\vbox{
\includegraphics[width=0.45\textwidth]{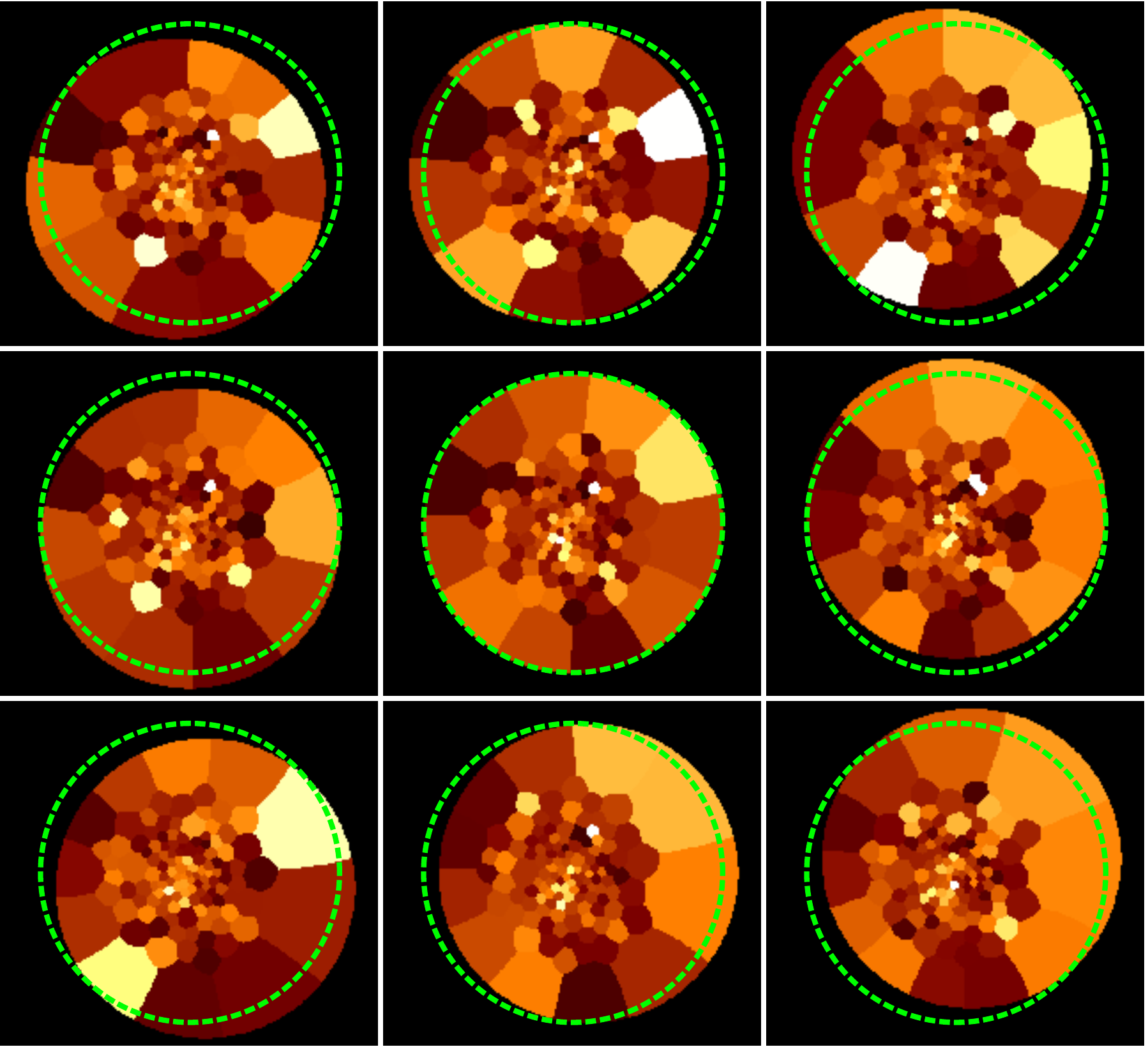}
\includegraphics[width=0.45\textwidth]{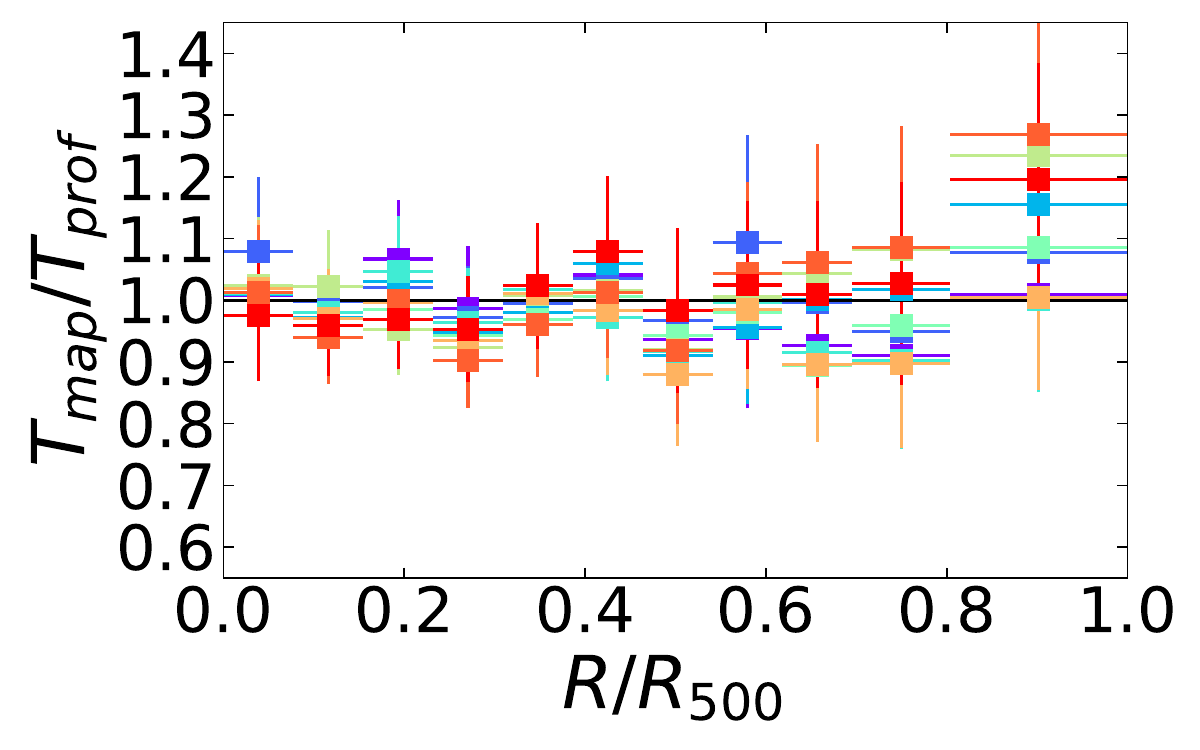}
}
\caption{
Impact of the Voronoi tessellation on the 2D temperature distribution. {\it Top:} temperature maps obtained for G041.45+29.10 by offsetting the center of the map wrt to the X-ray peak. The green circles identify $R_{500}$ centered on the peak. {\it Bottom:} ratio between the temperature recovered from the nine maps (using Eq. \ref{T1D}) and the temperature estimated using the spectra extracted from azimuthal annuli. Each color (from blue to red) represents one of the maps in the top panel (from top-left to bottom-right).
}
\label{fig:multimaps}
\end{figure}

\begin{figure}[h]
\centering
\includegraphics[width=0.5\textwidth]{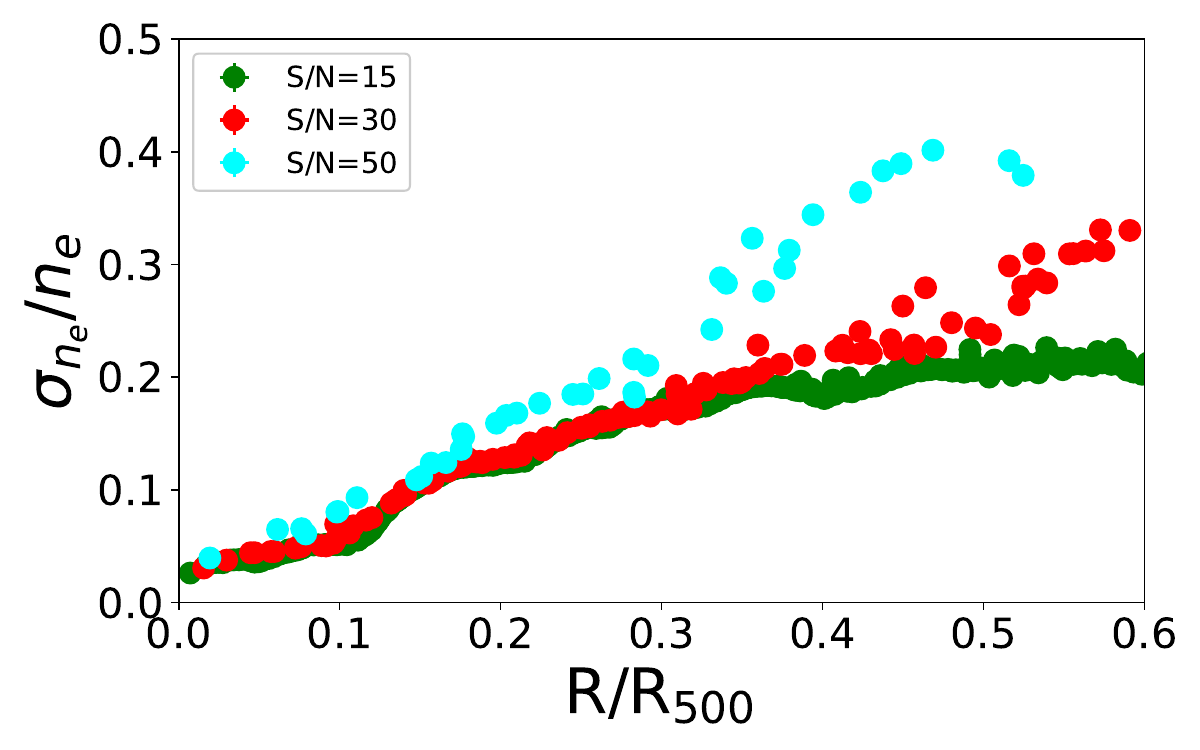}
\caption{
Impact of the map resolution on the density profile in the case of G041.45+29.10. We truncate the plot at 0.6$R_{500}$ because beyond that radius we have a poor resolution (in particular with S/N=50). Each point represents the value of $\sigma_{n_e}/n_e$ computed at the radius of the cells in the map.
}
\label{fig:SBres}
\end{figure}

Given a certain image, the binning mask obtained with WVT is deterministic (i.e., by running the algorithm multiple times, one obtains always the same result). 
Depending of the resolution of the mask, it is possible that some inhomogeneities are missed because they are spanning multiple cells, with the net result that the feature is washed out. 
In order to investigate how the choice of the map center impacts the obtained map and our results, we derived maps centered offset from the X-ray peak. 
The center of the new maps was computed from a grid with the X values obtained as $X_{center}$=$X_{peak}$+C$\times$R$_{500}$ and the Y values as $Y_{center}$=$Y_{peak}$+C$\times$R$_{500}$, where C can assume 3 values: -0.1, 0, +0.1. 
As an example, in the top panel of Fig. \ref{fig:multimaps}, we show the results for G041.45+29.10. Despite the different binning, the same features are observed in all the maps. 
In the bottom panel we show that the recovered profiles from the individual maps differ only by a few percent (and are always consistent within the uncertainties) as far as the map resolution is good enough (typically within $\sim$0.6R$_{500}$). 
At large radii, when the binning becomes quite coarse there is an increase of the scatter, although the data points are still consistent within the 1$\sigma$ errors. 
Similar results are obtained with the other clusters in the sample, so that the results of the paper are not affected by the deterministic nature of the Voronoi binning.

The results presented in the current analysis are based on the binning of temperature maps obtained with S/N=30. Such a choice is set by the requirement to have typical temperature uncertainties of the order of 10-20$\%$. 
However, the resolution of the maps can hide some relevant features and reduce or increase the observed scatter, in particular in the electron density maps used to infer the type of fluctuations. 
This is probably due to the increasing size of the binning in the outer regions, amplified by the presence of steep surface brightness gradients.
To verify that, we produced SB maps with S/N=15, 30, and 50 and we derived the scatter as a function of radius. 
In Fig. \ref{fig:SBres}  we show the results with the electron densities calculated under the assumption of $n_e=\sqrt{SB}$. 
Indeed, the ratio  $\sigma_{n_e}/n_e$ increases from low to high S/N.
However, it is worth noticing that the trend remains the same with an increase as a function of the radius (as discussed in Sect. \ref{sec:type} and shown in Fig. \ref{fig:scatterR}), although there is a flattening beyond 0.3-0.4$R_{500}$.

\section{Projection effects to the density and temperature perturbations }\label{2D3Deff}
\begin{figure}[h]
\centering
\includegraphics[width=0.5\textwidth]{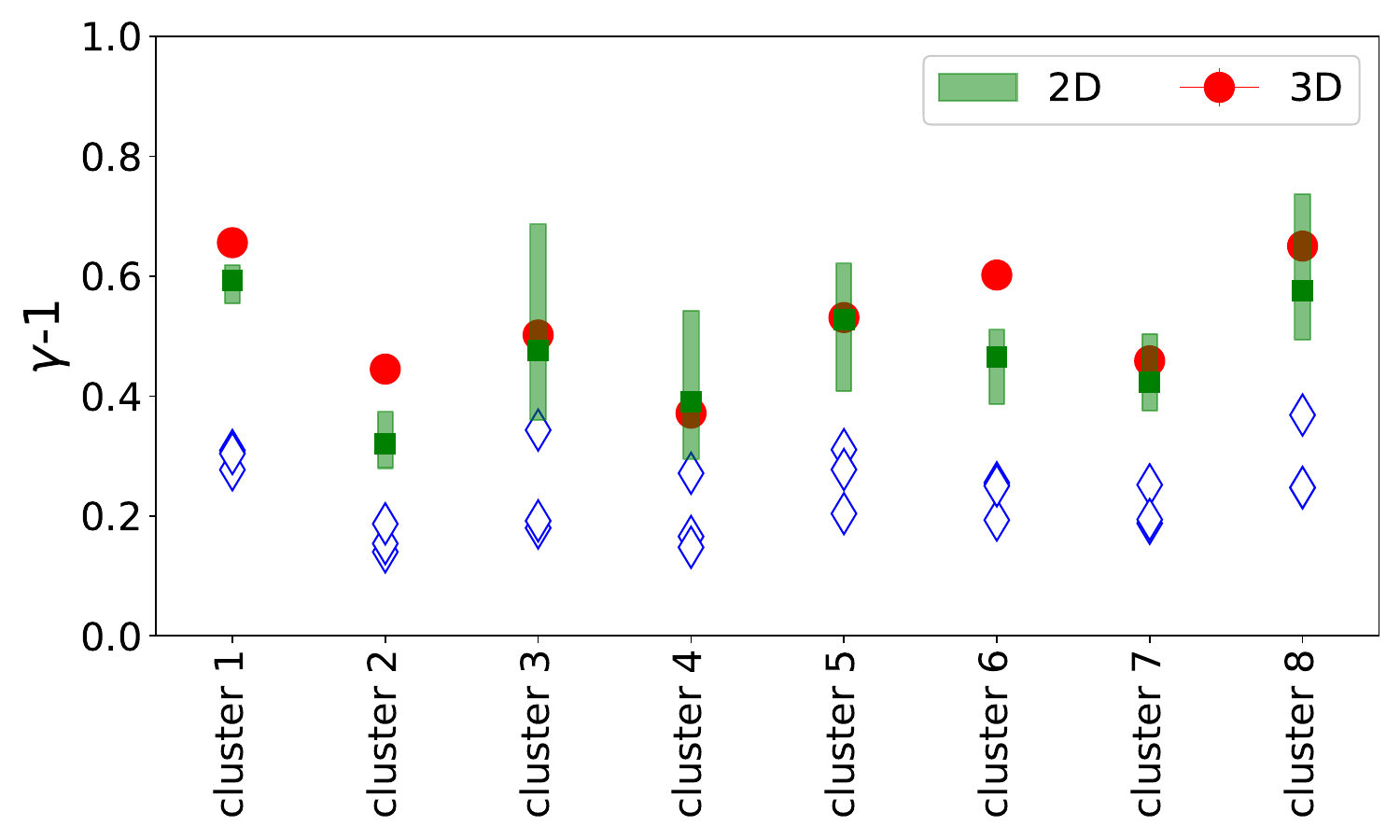}
\caption{
For each of the 8 simulated clusters, we show the mean ratio between the radial values of $\delta T_{3D}/T_{3D}$ and $\delta n / n$ (red points), $\delta T_{2D}/T_{2D}$ and $\delta n_{2D}^2 / n_{2D}^2$ in three different projections (blue diamonds), and the latter values corrected by
the factor of 2 from \cite{sch04} (the green squares represent the average values while the error bars represent the
minimum and maximum values from the different projections). 
}
\label{fig:projeff}
\end{figure}

By analyzing a set of 3D cosmological simulations (\citealt{vaz2017,simonte2022})  we tested how well Eq. \ref{eq:fluc2D} (derived following \citealt{sch04}) provides a good representation of the correlation between the projected density and temperature fluctuations (i.e., the relation of the 2D perturbations being a factor of $\sim$2 flatter than the relation between 3D density and temperature perturbations).
The simulations are performed with the code ENZO (\citealt{2014ApJS..211...19B}) and accurately capture the dynamical evolution of the ICM as galaxy clusters undergo matter accretion throughout their evolutionary process. 
Although the employed simulations are classified as non-radiative, previous studies have indicated that the influence of non-gravitational effects on the larger scales of interest (> few tenths of kpc) is comparatively limited, in contrast to the significant impact of mergers and accretion phenomena in shaping the properties of ICM (e.g., \citealt{vaz2012,2019ApJ...874...42V}).  
We refer the reader to \citet[][and references therein]{simonte2022} for further details. 
The 3D ICM perturbations are estimated in spherical shells with a size of 0.1$R_{500}$. 
In Fig. \ref{fig:projeff} we show as red points the fitted slope (i.e., $\gamma$-1) between the 3D density and temperature perturbations and as blue diamonds the 2D slopes from three different projections. 
It is clear that projection effects tend to flatten the relation as expected. 
The flattening is roughly a factor of 2 as derived following \cite{sch04} (see the green squares).

\end{appendix}

\end{document}